\documentclass[11pt]{article}
\usepackage{graphicx}
\usepackage{natbib} 
\usepackage{float}
\usepackage{caption}
\usepackage{subcaption} 
\usepackage{url} 
\usepackage{multirow}
\usepackage{makecell}
\usepackage{comment}
\usepackage{soul}
\usepackage{float}
\newcommand{\blind}{0}
\usepackage{authblk,amsmath,amssymb,graphicx,natbib}
\DeclareMathOperator{\Var}{Var}
\usepackage[margin=1in, letterpaper]{geometry}
\usepackage{setspace}
\usepackage[ruled,vlined]{algorithm2e}
\SetKwInput{Input}{Input}
\SetKwInput{Output}{Output}
\DontPrintSemicolon
\LinesNotNumbered             
\SetAlgoInsideSkip{smallskip} 
\setcitestyle{semicolon}

\begin{document}

\if0\blind
{
  \title{\bf Scalable Heteroskedastic Gaussian Process Models \\for Large Inhomogeneous Datasets}

\author[1,2]{K. Potter}
\author[1,2]{K.R. Moran}
\author[2]{R. Ulrich}
\author[1]{D.C. Stenning}
\author[1]{D. Bingham}
\author[3]{L. Castro}
\author[4]{G. Wilson}
\author[4]{C.A. Maldonado}

\affil[1]{Department of Statistics and Actuarial Science, Simon Fraser University}
\affil[2]{Statistics Group (CAI-4), Los Alamos National Laboratory}
\affil[3]{Information Systems and Modeling (A-1), Los Alamos National Laboratory}
\affil[4]{Space Science and Applications (ISR-1), Los Alamos National Laboratory}

  \maketitle
} \fi

\if1\blind
{
  \bigskip
  \bigskip
  \bigskip
  \begin{center}
    {\LARGE\bf Title}
\end{center}
  \medskip
} \fi

\bigskip
\begin{abstract}

We introduce Heteroskedastic Normalized Vecchia Gaussian Processes (HetNV), a scalable framework for Gaussian process regression with input-dependent observation noise. HetNV combines Vecchia likelihood approximations on normalized inputs with residual-based nonparametric variance estimation. The latent mean is estimated via a Vecchia GP with observation-specific nugget variances, while the log noise variance is obtained by smoothing stabilized log-squared residual pseudo-responses that account for kriging uncertainty and current nugget estimates, using LOESS in one dimension and thin plate spline generalized additive models in two dimensions. The method alternates between mean and variance updates, avoiding latent-variable inference for the variance process. For fixed neighborhood size ($m$) and number of observations ($n$), the dominant per-iteration cost is the Vecchia update, scaling as \(O(nm^2)\). Simulation studies show improved recovery of input-dependent uncertainty relative to homoskedastic Vecchia models while maintaining competitive mean prediction accuracy, with additional gains in two-dimensional mean estimation. An application to spacecraft plasma measurements demonstrates how locally adaptive uncertainty estimates influence downstream signal-detection decisions in large, noisy, heteroskedastic settings.

\end{abstract}

\newenvironment{keywords}{
  \vspace{2mm}
  \noindent\textbf{Keywords:}
}{\par\vspace{2mm}}
\begin{keywords}
Gaussian Processes; Heteroskedasticity; Scalable Inference; Uncertainty Quantification; Vecchia Approximation
\end{keywords}

\vfill

% \title{}
% \author{\vspace{-1ex} K. Potter, K. Moran, D. Stenning, D. Bingham, R. Ulrich, L. Castro, C. Maldonado}
% \date{}
% \maketitle

\section{Introduction}

Gaussian processes (GPs) provide flexible and interpretable nonparametric models for complex functions, with applications commonly arising in spatial statistics and computer experiments. They produce nonparametric function estimates and support coherent uncertainty quantification \citep{rasmussen06,stein99}. In geostatistics, GPs correspond to classical kriging models for spatial fields \citep{cressie11}; in computer experiments, they are widely used as surrogate models for complex simulators (e.g., \citealp{Sacks1989,kennedy01,gramacy20}). A GP defines a distribution over functions $f(x)$, where $x \in \mathcal{X} \subset \mathbb{R}^d$ denotes an input (covariate) location, such that any finite collection of function values follows a multivariate normal distribution. Given observations $i=1,\ldots,n$ observations of $y_i = f(x_i) + \varepsilon_i$ with $\varepsilon_i \sim N(0,\sigma^2)$, standard GP regression assumes independent homoskedastic noise with constant variance $\sigma^2$.

% Given observations $y_i = f(x_i) + \varepsilon_i$ at input locations $x_i \in \mathcal{X}$, with independent noise $\varepsilon_i \sim N(0, \sigma^2)$, standard GP regression yields closed-form predictive distributions for $f(x)$ at new input locations $x$ \citep{rasmussen06}.

Two core limitations restrict the applicability of GPs in modern settings. First, exact inference for GPs scales as $\mathcal{O}(n^3)$ due to repeated Cholesky decompositions of $n \times n$ covariance matrices \citep{rasmussen06}. This limits the use of GPs beyond moderate sample sizes. Second, the homoskedastic noise assumption is often unrealistic. In many real systems, noise levels depend on operating conditions, local dynamics, or measurement context. Heteroskedasticity arises, for example, in sensor measurements with varying signal-to-noise ratios, simulators with input-dependent fidelity, and environmental monitoring with changing background conditions (e.g., \citealp{Goldberg1997,Binois2018}). Neglecting input-dependent noise can lead to under- or over-dispersed predictive intervals and miscalibrated uncertainty \citep{Goldberg1997,Binois2018}.

A large body of work has focused on improving the scalability of Gaussian process inference. Existing approaches include inducing-point and sparse variational methods \citep{quinonero05}, low-rank approximations \citep{cressie08}, covariance tapering \citep{kaufman08}, multi-resolution representations \citep{nychka15}, and mixtures of local GPs \citep{gramacy15}. Within this landscape, Vecchia approximations \citep{vecchia88} provide a flexible framework for large-scale GP regression by replacing the full joint density of the observation $\mathbf{y} = (y_1,\dots,y_n)$ with a product of low-dimensional conditional densities. Each observation conditions only on a small set of previously ordered observations, inducing a sparse precision matrix and reducing computational cost to $\mathcal{O}(nm^2)$ for a neighborhood size $m \ll n$ \citep{stein04,guinness18,katzfuss21}. %Extensions of Vecchia \citep{vecchia88} have explored learning anisotropic input scalings so that neighborhoods are defined in an estimated correlation geometry, improving likelihood approximation and prediction \citep{katzfuss22}.

Heteroskedastic Gaussian process models generalize standard GPs by allowing the noise variance to depend on the input, $\mathrm{Var}(\varepsilon_i) = \sigma^2(x_i)$. This is important for stochastic simulators, observational data, and scientific measurement systems in which variability changes with operating conditions, local dynamics, or signal strength. Classical GP surrogates for deterministic computer experiments are often used as interpolators with a small or zero-valued nugget term \citep{Sacks1989,gramacy20,GramacyLee2012}, but stochastic simulators and measurement processes instead motivate models of the form $y(x) = f(x) + \varepsilon(x)$ with $\mathrm{Var}(\varepsilon(x)) = \sigma^2(x)$ \citep{Ankenman2010,Binois2018}. Existing heteroskedastic GP approaches include latent log-variance processes \citep{Goldberg1997}, variational approximations \citep{lazaro11}, convex reformulations \citep{le05}, and replication-based methods \citep{Binois2018}. However, many of these approaches rely on dense computations, replicated designs, or both, limiting their applicability to large datasets without replicated inputs.

What remains missing is a method that is simultaneously scalable, heteroskedastic, and applicable without replicated inputs. To address this gap we introduce \emph{Heteroskedastic Normalized Vecchia Gaussian Process} regression (HetNV). Unlike existing heteroskedastic GP approaches that rely on latent Gaussian processes for the variance function or require replicated inputs, HetNV avoids latent-variable inference and instead estimates the variance surface using stabilized residuals combined with nonparametric smoothing. This design enables scalable inference through the Vecchia approximation while retaining the ability to model input-dependent noise in large, unreplicated datasets.

HetNV constructs the Vecchia approximation on inputs normalized to the unit cube $[0,1]^d$, placing all input dimensions on a comparable scale for neighborhood selection without otherwise modifying the GP model. The mean function is estimated using a Vecchia likelihood, while the noise variance function $\sigma^2(x)$ is estimated from smoothed, variance-stabilized residuals. In one dimension, we model $\log \sigma^2(x)$ using LOESS \citep{Cleveland1988}; in two dimensions, we use thin plate spline generalized additive models (GAMs) \citep{wood2017}. %Stabilized residuals incorporate both uncertainty in the mean response induced by the GP and the current observation-level variance estimate, and a bias correction based on the $\chi^2_1$ distribution ensures that the resulting estimator targets the log of the noise variance. 
The mean and variance models are updated alternately in a plug-in scheme, avoiding latent-variable inference.

We validate HetNV on synthetic one- and two-dimensional heteroskedastic examples spanning a range of Matérn smoothness parameters and sample sizes. Across these settings, HetNV consistently improves uncertainty calibration relative to homoskedastic standardized Vecchia models, while maintaining comparable predictive accuracy and near-linear runtime in $n$. Where feasible, we compare against full heteroskedastic GP models with latent log-variance processes; HetNV achieves comparable predictive performance while scaling to substantially larger datasets at a fraction of the computational cost. Finally, we present an application of the method to spacecraft plasma measurements, where HetNV underpins an automated pipeline for denoising and peak extraction in time--energy spectrograms.

This paper makes three contributions. First, we introduce Heteroskedastic Normalized Vecchia Gaussian Processes (HetNV), a scalable framework for Gaussian process regression with input-dependent noise that combines Vecchia likelihood approximations with residual-based variance estimation, avoiding latent-variable modeling. 
Second, we show that explicitly modeling heteroskedasticity improves uncertainty calibration and, in multi-dimensional settings, can improve mean-surface estimation through adaptive reweighting of observations. 
Third, we demonstrate that the proposed method is practical for large datasets through simulation studies and a real-data application to spacecraft plasma measurements, where uncertainty estimates directly influence downstream signal-detection decisions.
Computationally, the key feature of HetNV is that input-dependent noise is introduced through smooth residual-based pseudo-responses rather than through a second latent GP, preserving sparse Vecchia computations throughout. %From a computational-statistics perspective, the key feature of HetNV is that heteroskedasticity is introduced without requiring a second latent GP or dense covariance operations. The method instead combines sparse Vecchia likelihood updates with low-dimensional smoothers for stabilized residual pseudo-responses. This yields an implementation that can be fit with standard GP and smoothing tools, produces interpretable variance surfaces, and remains practical in settings where fully latent heteroskedastic GP models become computationally expensive.

The remainder of the paper is organized as follows. Sections~\ref{sec:bg_vec} and \ref{sec:bg_het} review Vecchia approximations and existing heteroskedastic GP models. Section~\ref{sec:method} introduces the HetNV methodology, emphasizing the alternating mean–variance estimation scheme. Section~\ref{sec:experiments} presents simulation results in one and two dimensions. Section~\ref{sec:aps} describes an application to space plasma measurements. Section~\ref{sec:dis} discusses practical considerations and limitations, and Section~\ref{sec:conclusion} concludes and outlines directions for future work.

\subsection{Vecchia Approximations for Scalable Gaussian Processes}\label{sec:bg_vec}

The Vecchia approximation, introduced by \citet{vecchia88} in spatial statistics and later extended to GP regression \citep{stein04,guinness18,katzfuss21}, approximates the joint density of $\mathbf{y}$ as:
% \begin{equation}
  \[  p(y_1,\dots,y_n) \approx \prod_{i=1}^n p(y_i \mid y_{\mathcal{N}(i)}),\]
    % \label{eq:vecchia-factor}
% \end{equation}
where $\mathcal{N}(i) \subset \{1, \dots, i-1\}$ is a chosen neighborhood for observation $i$. When each neighborhood is of size $m \ll n$, the resulting sparse precision matrix reduces Cholesky computations to $\mathcal{O}(nm^2)$, making the method suitable for large-scale problems. This approximation becomes exact as $m \to n-1$, and in practice even modest neighborhood sizes yield accurate inference when neighborhoods are chosen to reflect the correlation structure induced by the GP covariance function over the inputs \citep{katzfuss21}.

From a likelihood perspective, the Vecchia approximation can be interpreted as a form of conditional composite likelihood, constructed as a product of low-dimensional conditional densities \citep{Lindsay1988,Varin2011}. This connects Vecchia methods to a broader class of composite likelihood and pseudolikelihood approaches, where inference is based on tractable local components rather than the full joint likelihood \citep{Besag1974}. Unlike general composite likelihood constructions, however, the Vecchia approximation still defines a valid joint Gaussian distribution with an associated sparse precision matrix, enabling efficient inference and prediction in large-scale settings.

Ordering and conditioning-set selection play a central role in the accuracy of Vecchia approximations. Simple heuristics such as coordinate sorting often yield poor performance, while maximin ordering or space-filling curves improve approximation quality by encouraging spatial coverage \citep{guinness18,katzfuss21}. Recent work has also explored grouping strategies that reduce computational overhead while preserving accuracy \citep{guinness18,katzfuss21}. These choices affect both predictive performance and the quality of likelihood approximation.

Several extensions of this framework aim to improve approximation quality by adapting the geometry used to define neighborhoods. In particular, the scaled Vecchia approximation of \citet{katzfuss22} learns anisotropic input scalings by maximizing an approximate likelihood over scale parameters and other covariance hyperparameters, effectively estimating a correlation geometry in which neighborhoods are selected. By defining neighborhoods in this learned geometry, scaled Vecchia can improve parameter estimation and prediction while retaining near-linear computational complexity.

In contrast, the approach developed in this paper employs a normalized Vecchia approximation, in which each input dimension is rescaled to the unit interval prior to neighborhood selection and likelihood approximation. Rescaling inputs to the unit cube provides a simple and numerically stable reference scale for neighborhood selection and kernel evaluation, ensuring comparable distances across dimensions without introducing additional parameters \citep{rasmussen06,stein99,guinness18,katzfuss21}. This approach does not involve learning anisotropic scaling parameters or modifying the correlation geometry. As a result, it may be less adaptive than scaled Vecchia approaches that estimate an anisotropic correlation metric \citep{katzfuss22}, particularly when the true process exhibits strong directional dependence. We adopt fixed standardization here to reduce optimization complexity and to preserve computational simplicity in the heteroskedastic setting.

%Other scalable GP frameworks include inducing-point methods \citep{quinonero05}, low-rank approximations \citep{cressie08}, covariance tapering \citep{kaufman08}, and multi-resolution or local approximations \citep{nychka15,gramacy15,datta16}. 
Many other scalable GP approaches restrict spatial dependence either explicitly or implicitly. Vecchia approximations offer greater flexibility by accommodating both local and long-range dependence through neighborhood design. Recent theoretical results establish consistency and asymptotic normality of Vecchia-based estimators under increasing-domain asymptotics when neighborhood sizes grow appropriately \citep{guinness18,katzfuss21}. These properties motivate our use of Vecchia approximations in large-scale heteroskedastic settings.

\subsection{Heteroskedastic Gaussian Process Regression}\label{sec:bg_het}
Traditional GP models assume constant observation noise, $\mathrm{Var}(\varepsilon_i) = \sigma^2$. Heteroskedastic extensions relax this assumption by allowing $\mathrm{Var}(\varepsilon_i) = \sigma^2(x_i)$, which is important when uncertainty depends on system dynamics, measurement conditions, or local signal strength. Ignoring such structure can lead to miscalibrated predictive intervals and overconfident inference \citep{Goldberg1997,Binois2018,caldarelli23}. %Traditional GP models assume constant observation noise, $\mathrm{Var}(\varepsilon_i) = \sigma^2$. Heteroskedastic extensions relax this assumption by allowing the noise variance to depend on the input, $\mathrm{Var}(\varepsilon_i) = \sigma^2(x_i)$. This generalization is important in settings where uncertainty depends on system dynamics or measurement conditions. For example, in simulation experiments, variance often increases in specific regions of the input space. Ignoring heteroskedasticity can lead to miscalibrated predictive intervals and overconfident inference \citep{Goldberg1997,Binois2018,caldarelli23}.

A common approach is to place GP priors on both the latent mean function and the log-variance function \citep{Goldberg1997}. This hierarchical formulation is conceptually appealing but requires integration over two latent processes and is computationally expensive. Variational approximations and expectation propagation improve tractability \citep{lazaro11}, but typically rely on dense matrix operations and therefore scale poorly to large datasets. %Several modeling strategies have been proposed for heteroskedastic GP regression. The hierarchical approach introduced by \citet{Goldberg1997} assigns independent GP priors to the latent mean function and the log-variance function. As mentioned above, while conceptually appealing, inference requires integrating over two latent processes and is computationally expensive. Variational approximations and expectation propagation have been developed to improve tractability \citep{lazaro11}, but these methods typically rely on dense matrix operations and do not scale well to large datasets.

Replication-based methods provide another route to heteroskedastic modeling. In stochastic simulation settings, repeated observations at common design sites enable direct estimation of local noise variance and support sequential design strategies that balance exploration and variance reduction \citep{Ankenman2010,Binois2018}. However, these computational advantages rely on shared inputs and diminish when observations are unique, as is common in spatial, temporal, and observational datasets. %Many heteroskedastic GP approaches are naturally formulated as hierarchical models, placing a GP prior on both the mean and the log-variance processes \citep{Goldberg1997}. These formulations are closely tied to stochastic simulation settings, where replication enables direct estimation of local noise and informs sequential design strategies that allocate computational effort between exploration and variance reduction \citep{Ankenman2010,Binois2018}. In practice, however, replicated designs are often unavailable in spatial and observational datasets, motivating alternative approaches that can recover input-dependent variance structure without explicit replication.

Other approaches reformulate heteroskedastic GP regression as a deterministic optimization problem. For instance, \citet{le05} cast variance estimation as convex regression on natural exponential family parameters, computing a maximum a posteriori (MAP) estimate via Newton’s method. Hard EM-like procedures that iteratively update the latent response and input-dependent noise processes have been proposed \citep{Kersting2007}, while related methods jointly optimize the mean- and noise-process hyperparameters \citep{quadrianto09}. These approaches avoid full latent-variable integration by relying on optimization-based approximations, although uncertainty quantification typically relies on additional approximations around the optimized solution. Computationally, they generally retain cubic complexity in $n$ due to dense covariance matrix operations.

Some methods exploit replicated designs, where repeated observations are available at identical input locations (or design sites), enabling direct estimation of local noise variance (e.g., \citealp{Ankenman2010}). Approaches such as stochastic kriging \citep{Ankenman2010} and the heteroskedastic GP framework of \citet{Binois2018} build on this idea by introducing latent variance parameters for each unique input location (i.e., each design site) and imposing smoothness through a GP prior on the log-variance. The computational advantages of these methods rely on shared inputs and diminish when observations are unique, as is common in spatial and temporal applications.

An alternative class of methods estimates the variance function using residuals from an initial GP mean fit \citep[e.g.,][]{Goldberg1997,Kersting2007,Binois2018}. In this two-stage approach, transformed log-squared residuals are modeled using nonparametric regression tools such as generalized additive models (GAMs). Although these methods do not perform full joint Bayesian inference over the variance process, they are computationally efficient and can still provide approximate uncertainty quantification for the estimated variance surface and well-calibrated predictive intervals in practice. This perspective is closely related to the GAMLSS framework of \citet{rigby05}, which allows distributional parameters such as scale to vary flexibly with covariates. Existing heteroskedastic GP methods generally perform well in low-sample or replicated-design settings, but do not scale efficiently to high-resolution spatial or spatiotemporal domains with complex, input-dependent noise. Our approach addresses this gap by combining a scalable likelihood approximation with simple, dimension-aware variance modeling, yielding a practical heteroskedastic GP framework for large datasets.

%The proposed approach is particularly useful in settings with large, unreplicated datasets and spatially or temporally varying noise, where standard homoskedastic GP models can lead to miscalibrated uncertainty and suboptimal inference. These settings arise frequently in spatial statistics, environmental monitoring, and scientific measurement systems with input-dependent signal-to-noise characteristics.

\section{Methodology: Heteroskedastic Normalized Vecchia Gaussian Processes}
\label{sec:method}

We model observations $y_i$ at input locations $x_i \in \mathcal{X} \subset \mathbb{R}^d$ as
\[
y_i = f(x_i) + \varepsilon_i, \qquad \varepsilon_i \sim N\big(0, \sigma^2(x_i)\big), \qquad i=1,\ldots,n,
\]
where $f(\cdot)$ is a latent mean function, $\varepsilon_i$ are independent noise terms, and $\sigma^2(\cdot)$ is an unknown, input-dependent noise variance function. HetNV estimates the mean function using a standardized Vecchia-approximated Gaussian process and estimates the variance function using residual-based nonparametric smoothing. We use “kriging variance” to refer to the posterior variance of the latent GP $f$.

Estimation proceeds via an alternating plug-in scheme, summarized in Algorithm~\ref{alg:hetsv}. The individual components of this procedure are described in detail in the subsections that follow.
    
\begin{algorithm}[H]
\caption{HetNV alternating mean--variance estimation}
\label{alg:hetsv}
\Input{Observations $\{(x_i,y_i)\}_{i=1}^n$; neighborhood size $m$; iterations $T$; stabilization $\alpha>0$; bias correction $\delta=-\mathbb{E}\{\log(\chi^2_1)\}$ $=$ $-\psi(1/2)-\log 2$ $\approx 1.2704$.}

\Output{Posterior summaries of the latent mean function $f(\cdot)$, including $\hat f(x)$ and $\widehat{\Var}\{f(x)\}$, together with the estimated variance surface $\hat\sigma^2(\cdot)$.}

% \textbf{Rescale inputs:} $\tilde x_i \leftarrow (x_i - x_{\min}) / (x_{\max} - x_{\min})$ (component-wise), so that $\tilde x_i \in [0,1]^d$.\;
\textbf{Normalize inputs:} first standardize each input dimension and then affinely rescale the standardized coordinates to obtain $\tilde x_i\in[0,1]^d$; use $\tilde x_i$ for Vecchia neighborhood selection, covariance evaluation, and variance smoothing.\;

\textbf{Initialize:} set $\hat\sigma^{2\,(0)}(x_i)\leftarrow \hat\sigma_0^2$ (e.g., the constant nugget from a homoskedastic fit).\;

\For{$t=1$ \KwTo $T$}{
   \textbf{Mean update (Vecchia GP):} fit the Vecchia GP on normalized inputs $\tilde x_i$ using nugget variances $\hat\sigma^{2\,(t-1)}(x_i)$ and compute $\hat f^{(t)}(x_i)$ and kriging variances $v_i^{(t)}=\widehat{\Var}\{f(x_i)\}$.\;   

  \textbf{Variance pseudo-response:} set residuals $r_i^{(t)} \leftarrow y_i-\hat f^{(t)}(x_i)$ and form
  $\tilde g_i^{(t)} \leftarrow \log\big((r_i^{(t)})^2 + \alpha\{v_i^{(t)}+\hat\sigma^{2\,(t-1)}(x_i)\}\big)+\delta$.\;

 \textbf{Smooth log-variance:} fit $h^{(t)}(\tilde x_i)$ to $\tilde g_i^{(t)}$ using LOESS (1D) or thin plate spline GAM (2D), then update
$\hat\sigma^{2\,(t)}(x_i) \leftarrow \exp\{\hat h^{(t)}(\tilde x_i)\}$.\;

  \textbf{Optional robustification:} truncate extreme $\tilde g_i^{(t)}$ and/or shrink $\hat\sigma^{2\,(t)}(x)$ toward a baseline in sparse regions.\;
}
\end{algorithm}

\vspace{0.3cm} %adding a bit of space after the alg. 

\subsection{Mean model: normalized Vecchia Gaussian process approximation}

We place a zero-mean Gaussian process prior on $f$,
\[
f \sim \mathcal{GP}\big(0, K_\theta(\cdot,\cdot)\big),
\]
where $K_\theta$ is a stationary covariance kernel with hyperparameters $\theta$, such as marginal variance, range, and smoothness. In all experiments we use Matérn kernels with smoothness $\nu \in \{1.5, 2.5, \infty\}$, though the methodology is not restricted to this choice.

To scale inference to large datasets, we employ the Vecchia approximation \citep{vecchia88,stein04,guinness18,katzfuss21}. As previously stated, we refer to inputs as \emph{normalized} when they have been transformed by first standardizing each dimension to zero mean and unit variance and then rescaling to the unit cube $[0,1]^d$. The normalization step is not intended as a new approximation in itself; rather, it provides a stable and reproducible geometry for applying Vecchia conditioning within the proposed heteroskedastic fitting scheme. This combined transformation improves numerical stability for covariance estimation while ensuring that distances used for neighborhood selection are comparable across dimensions. This definition of normalization differs from standard usage in machine learning, where normalization typically refers only to unit-cube scaling; here we explicitly include both standardization and unit-cube rescaling. Given an ordering of the observations, the Vecchia approximation is applied to the marginal observation density, with covariance evaluated on the normalized inputs $\tilde x_i$:
\[
p(y_1,\dots,y_n) \approx \prod_{i=1}^n p\big(y_i \mid y_{\mathcal{N}(i)}\big).
\]
% \label{eq:vecchia-factor2}
% \end{equation}
where $\mathcal{N}(i) \subset \{1,\dots,i-1\}$ is a conditioning set of size at most $m$. This approximation induces a sparse precision matrix and reduces computational cost to $\mathcal{O}(nm^2)$.
Observations are modeled as
\[
y_i \mid f_i \sim \mathcal{N}\big(f_i, \sigma^2(x_i)\big),
\]
where $\sigma^2(x_i)$ is an observation-specific nugget variance. We implement Vecchia inference using the \texttt{GPvecchia} package. This preprocessing separates statistical scaling from geometric scaling: standardization stabilizes covariance parameter estimation, while unit-cube normalization ensures meaningful distance calculations for nearest-neighbor selection. Unlike scaled Vecchia approaches \citep{katzfuss22}, no anisotropic scaling parameters are learned; scaling is fixed and applied only for numerical stability and neighborhood construction.

\subsection{Variance model: residual-based log-variance smoothing}

HetNV estimates an input-dependent noise variance by smoothing stabilized residual information from the current Vecchia mean fit. Since residual variation reflects both observation noise and imperfections in the fitted mean, especially in unreplicated data, the variance update is designed to recover a stable empirically reasonable noise surface rather than a pointwise estimate guaranteed to equal the unknown true variance function \citep{HallCarroll1989, WangBrownCaiLevine2008}.

\subsubsection{Residual-based variance estimation}
Given the current mean-model fit, let $\hat f(x_i)$ denote the posterior mean of the latent function at $x_i$ and let
\[v_i = \widehat{\mathrm{Var}}\{f(x_i)\}\]
denote the corresponding kriging variance. Define residuals
\[r_i = y_i - \hat f(x_i).\]

The variance update is motivated by classical residual-based, plug-in variance-function estimation. In heteroskedastic nonparametric regression, a common strategy is to first estimate the mean function, form residuals, and then smooth squared or transformed residuals to recover the conditional variance function (e.g., \citealp{HallCarroll1989}; \citealp{Ruppert1997}; \citealp{davidian87}, \citealp{FanYao1998}). This idea is attractive in the present setting because it avoids introducing a second latent Gaussian process for the variance while still allowing the noise level to vary smoothly over the input space.

If the latent mean function were known, then under the Gaussian observation model
\[
y_i = f(x_i) + \epsilon_i, 
\qquad 
\epsilon_i \sim N\{0,\sigma^2(x_i)\},
\]
the residual $r_i = y_i - f(x_i)$ would satisfy
\[
\frac{r_i}{\sigma(x_i)} \sim N(0,1).
\]
Therefore,
\[
\frac{r_i^2}{\sigma^2(x_i)} \sim \chi^2_1.
\]
Taking logarithms gives
\[
\log(r_i^2)
=
\log \sigma^2(x_i) + \log(\chi^2_1),
\]
and hence
\[
\mathbb{E}\{\log(r_i^2)\}
=
\log \sigma^2(x_i) + \mathbb{E}\{\log(\chi^2_1)\}.
\]
Thus, the log-squared residual is a noisy, biased observation of the log noise variance. This motivates the analytic bias correction
\[
\delta = -\mathbb{E}\{\log(\chi^2_1)\}
=
-\psi(1/2)-\log 2
\approx 1.2704,
\]
where $\psi(\cdot)$ is the digamma function. After this correction, the transformed residual approximately targets $\log \sigma^2(x_i)$.

A direct estimator of $\sigma^2(x_i)$ based on $r_i^2$ is highly variable, and the raw 
log-squared residual $\log(r_i^2)$ can be unstable when $r_i^2$ is close to zero. We therefore 
work with stabilized log-squared residual pseudo-responses. The use of the log scale is 
motivated by two considerations: it enforces positivity of the variance after exponentiation, 
and it converts multiplicative variance structure into an additive smoothing problem. This 
log-scale treatment is consistent with heteroskedastic GP models that place structure on an 
input-dependent variance or log-variance process \citep{Goldberg1997,lazaro11}, as well as 
distributional regression frameworks in which scale parameters are modeled as smooth functions 
of covariates \citep{rigby05}.

We define
\[
\tilde g_i
=
\log\big( r_i^2 + \alpha \{ v_i + \hat\sigma^2(x_i) \} \big) + \delta,
\]
where $\alpha > 0$ is a stabilization parameter and $\delta$ is a bias correction. The term 
$v_i$ accounts for posterior uncertainty in the fitted latent mean, while the current nugget 
estimate $\hat\sigma^2(x_i)$ provides a local estimate of the observation-level noise scale. 
Thus, $v_i+\hat\sigma^2(x_i)$ approximates the current local predictive uncertainty for an 
observation and acts as a data-adaptive regularizer inside the logarithm.

This construction combines several established ideas: residual-based variance-function estimation, log-scale modeling of positive variance functions, and analytic bias correction for log-squared Gaussian residuals. The GP-specific adaptation is the inclusion of the kriging variance and current nugget estimate in the stabilizing term. These additions regularize the variance update in large, unreplicated datasets, where residual variability may otherwise be difficult to separate from local mean-process interpolation error.

Because $r_i$ is computed from an estimated mean and because the stabilizing term modifies the exact log-squared-residual identity, $\tilde g_i$ should be interpreted as an approximate pseudo-response for the log variance rather than an unbiased observation of it. In implementation, extreme pseudo-responses may also be truncated to improve robustness against isolated large residuals or local approximation artifacts.

\subsubsection{Nonparametric variance smoothing}
We represent the noise variance through a smooth log-variance function
\[
\log \sigma^2(x_i) = h(\tilde{x}_i).
\]
where $h(\cdot)$ is unknown. Modeling variance on the log scale ensures positivity of 
$\sigma^2(x)$ after exponentiation while allowing $h(\tilde{x})$ to be estimated using unconstrained 
smoothing methods. This choice is consistent with heteroskedastic GP formulations that model 
input-dependent noise through a latent variance or log-variance process 
\citep{Goldberg1997,lazaro11}, and with distributional regression frameworks in which scale 
parameters vary smoothly with covariates \citep{rigby05}.

If residuals were computed using the true mean function and no stabilization were applied, the 
bias-corrected log-squared residuals would target $h(x_i)$ in expectation. In practice, 
$\tilde g_i$ is only an approximate pseudo-response for $h(\tilde{x_i})$ because residuals are computed 
from the fitted mean and include the stabilization term. We therefore estimate $h(\cdot)$ by 
smoothing the pseudo-responses $\tilde g_i$ over the input space, treating them as noisy, 
regularized measurements of the log-variance surface.

For one-dimensional inputs, we estimate $h(\tilde{x})$ using LOESS smoothing, following the local 
smoothing perspective used in nonparametric variance-function estimation 
\citep{Ruppert1997,FanYao1998}. For two-dimensional inputs, we estimate $h(\tilde{x_1},\tilde{x_2})$ using a 
thin plate spline GAM \citep{wood2017}. Thin plate spline GAMs provide a flexible penalized 
smooth surface without requiring a parametric form for the variance function, making them a 
natural smoother for the residual-derived pseudo-responses.

This smoother-based update is also related to generalized additive models for location, scale, 
and shape, where distributional parameters such as scale are modeled semiparametrically as 
smooth functions of covariates \citep{rigby05}. Relative to fully latent heteroskedastic GP 
formulations, the smoother-based update acts as an implicit regularizer by restricting the 
flexibility of the variance surface and reducing sensitivity to localized residual artifacts.

While thin plate spline GAMs can, in principle, be extended to higher-dimensional inputs, they 
become computationally demanding and less stable as dimension increases. Estimation for 
higher-dimensional inputs, not considered here, can instead be carried out using more scalable 
structured models, such as additive GAMs, tensor-product smoothers, or low-rank semiparametric 
representations \citep{ruppert2003,wood2006,wood2017}. During smoothing, observations can be 
weighted using a combination of inverse predictive variance and robust residual-based weights, 
downweighting locations with high uncertainty or extreme pseudo-responses.

\subsubsection{Robustification and shrinkage}

Isolated large residuals can lead to unstable local variance estimates, particularly in sparsely sampled regions. Such behavior is a known challenge for residual-based variance estimation because squared 
residuals can be strongly influenced by outliers, local mean-estimation error, or sparse-design 
artifacts \citep{HallCarroll1989,Ruppert1997}. To mitigate this effect, we shrink each variance update toward the previous iteration's nugget surface on the log scale. Let $\tilde{\sigma}^{2(t)}(x)$ denote the variance estimate obtained by smoothing the pseudo-responses at iteration $t$. We update
\[
\log \hat{\sigma}^{2(t)}(x)
=
(1-\lambda)\log \hat{\sigma}^{2(t-1)}(x)
+
\lambda \log \tilde{\sigma}^{2(t)}(x),
\]
where $\lambda \in (0,1]$ controls the strength of the update. The update is performed on the log scale because the variance model is defined through 
$h(\tilde{x})=\log\sigma^2(x)$. Shrinkage on this scale corresponds to geometric averaging of variance 
estimates, preserves positivity after exponentiation, and prevents a single smoothed update from dominating the nugget surface. Smaller values of $\lambda$ yield stronger shrinkage toward the previous nugget estimate, while $\lambda=1$ corresponds to no shrinkage. After this update, extreme values are truncated to fixed lower and upper quantiles of the updated log-variance surface to prevent isolated residuals from producing unstable nugget estimates. This step improves numerical robustness while preserving large-scale heteroskedastic structure.

\subsection{Alternating plug-in estimation}

Mean and variance estimation in HetNV proceeds via the following alternating scheme for $t = 1,\ldots, T$:
\begin{enumerate}
\item \textbf{Mean update:} Given the current variance estimates $\hat\sigma^{2\,(t-1)}(x_i)$, estimate GP hyperparameters $\hat{\theta}$ by maximizing the Vecchia log-likelihood, and compute updated mean predictions $\hat f^{(t)}(x_i)$ and kriging variances $v_i^{(t)}$.

\item \textbf{Variance update:} Given $\hat f^{(t)}(x_i)$ and $v_i^{(t)}$, form pseudo-responses $\tilde g_i^{(t)}$ and smooth to obtain an updated variance estimate $\hat\sigma^{2\,(t)}(x)$.
\end{enumerate}
The procedure is initialized with $\hat\sigma^{2\,(0)}(x_i)$, typically taken as the constant nugget estimate from a homoskedastic fit. The procedure is repeated for a small number of iterations $T$, typically between three and ten depending on sample size, with optional early stopping based on convergence of the variance estimates. In practice, we also monitor convergence of the alternating updates and terminate early when successive updates to the variance surface $\hat\sigma^2(x)$ (or equivalently the smoothed log-variance function) change by less than a specified tolerance.

The alternating updates are best interpreted as a plug-in estimation procedure rather than joint maximization of a single latent heteroskedastic GP likelihood. This is analogous to residual-based plug-in estimators for conditional variance functions, where the variance model is fit after substituting an estimated mean function into the residuals \citep{davidian87, ChiouMuller1999, YuPeace2012, Ruppert1997}. Here, ``plug-in'' means that each update treats the output of the other update as fixed: the mean update conditions on the current estimate of the variance surface, and the variance update is computed from residuals based on the current estimate of the mean function. This is analogous to residual-based plug-in estimators for conditional variance functions, where the variance model is fit after substituting an estimated mean function into the residuals \citep{HallCarroll1989,Ruppert1997,FanYao1998}.This distinction is deliberate. Fully latent heteroskedastic GP models introduce a second latent process for the log variance, yielding a coherent joint hierarchical model but substantially increasing computational and identifiability challenges, particularly in the absence of replicated inputs (e.g., \citealp{Goldberg1997}; \citealp{lazaro11}). In replicated stochastic computer experiments, within-site variability provides direct information about local noise variance and helps stabilize latent variance-process estimation (e.g., \citealp{Ankenman2010}; \citealp{Binois2018}). In contrast, our target setting consists primarily of large spatial or temporal datasets with essentially unique input locations, where local variability can often be explained either through latent mean-process structure or through increased observation noise.

In preliminary experiments with coupled latent mean and variance Vecchia GPs, we found that the variance process could become weakly identifiable relative to the mean process and overly sensitive to local neighborhood-level fluctuations. This issue is particularly pronounced in sparse nearest-neighbor approximations because highly local conditioning structures can amplify small interpolation artifacts in the residual process. HetNV therefore replaces the second latent GP with a semiparametric residual-smoothing update. The resulting variance model is intentionally less flexible than a second GP and consequently acts as an implicit regularizer on the heteroskedastic component.

The plug-in nature of HetNV means that the resulting predictive intervals do not fully account for uncertainty from all stages of estimation \citep{gelman2013bayesian,yang2015variance, ChiouMuller1999, YuPeace2012}. Unlike Gibbs sampling or other fully Bayesian approaches, which propagate posterior uncertainty through iterative sampling from conditional distributions, HetNV alternates between point estimates of the mean function and the noise variance surface. In addition, each mean update uses maximum likelihood estimates $\hat\theta$ of the GP hyperparameters, so hyperparameter uncertainty is not propagated. Consequently, the reported predictive variances should be interpreted as conditional on the fitted hyperparameters and estimated variance surface, and may underrepresent total uncertainty.

Similar computational tradeoffs appear in optimization- and hard-EM-like heteroskedastic GP procedures that replace full joint posterior inference with point estimation, including joint optimization and alternating estimation of the mean and variance processes; see, for example, \citealp{le05}, \citealp{Kersting2007}, and \citealp{quadrianto09}. In HetNV, each mean update remains likelihood-based conditional on the current nugget surface, while the variance update is performed using stabilized residual pseudo-responses and low-dimensional smoothing. The resulting procedure does not fully propagate uncertainty in the estimated variance surface into the final predictive distribution, so predictive intervals should be interpreted as plug-in intervals. However, this approximation substantially improves computational scalability and empirical stability in large unreplicated datasets while retaining flexible input-dependent uncertainty estimation.

To assess convergence of the alternating updates, we monitor the relative change in the estimated variance surface between successive iterations. Specifically, we compute the normalized $\ell_2$ difference between $\hat\sigma^{2\,(t)}$ and $\hat\sigma^{2\,(t-1)}$. In all simulation settings considered here, this quantity decreases rapidly and stabilizes within a small number of iterations (typically fewer than 5), indicating that the alternating scheme converges to a stable solution in practice.

\subsection{Prediction}

Predictive summaries for observations can be obtained by
\[
\hat y(x) = \hat f(x), \qquad
\widehat{\Var}\{y(x)\} = \widehat{\Var}\{f(x)\} + \hat\sigma^2(x).
\]
For a new input $x^\star$, let $\hat f(x^\star)$ and $\widehat{\mathrm{Var}}\{f(x^\star)\}$ denote the Vecchia-based posterior mean and variance of the latent function. For prediction, the new input $x^\star$ is transformed using the same normalization map as the training inputs to obtain $\tilde x^\star$, and the variance smoother is evaluated as $\hat\sigma^2(x^\star)=\exp\{\hat h(\tilde x^\star)\}$.The predictive variance for a new observation is
\[
\widehat{\mathrm{Var}}(y^\star \mid x^\star)
=
\widehat{\mathrm{Var}}\{f(x^\star)\} + \widehat{\sigma}^2(x^\star).
\]
We refer to this sum as the \emph{total predictive variance} and use it for coverage diagnostics in Section~\ref{sec:experiments}. Prediction intervals are constructed from the corresponding Gaussian predictive distribution,
\[
y^\star \mid x^\star \sim N\!\left(\hat f(x^\star),\ \widehat{\Var}\{f(x^\star)\} + \hat\sigma^2(x^\star)\right).
\]

\section{Experiments}
\label{sec:experiments}

We benchmark HetNV in two primary settings: one-dimensional heteroskedastic
regression and two-dimensional heteroskedastic regression. Across all cases we
sweep Matérn smoothness $\nu\in\{1.5,2.5,\infty\}$, with $\nu{=}\infty$ denoting
the squared-exponential kernel, and consider multiple problem sizes (or grid
resolutions in two dimensions). Each configuration is repeated over $100$ Monte Carlo (MC) replicates. Within each scenario, simulation parameters are held fixed and all methods are evaluated on identical data realizations in each replicate, so differences in performance reflect modeling choices rather than variability in the data-generating process. We report mean $\pm$ MC standard deviation of the root mean squared error (RMSE), empirical 95\% coverage at the observation level, and runtime in the main paper.

\paragraph{Competitors:}
HetNV is our proposed heteroskedastic standardized Vecchia GP with residual-based variance estimation, as described in Section~\ref{sec:method}. 
HomoSV denotes the corresponding homoskedastic standardized Vecchia GP with a constant noise variance. Both Vecchia-based models were implemented using the \texttt{GPvecchia} R package. For comparison, heteroskedastic GP models were fit using the \texttt{hetGP} R package \citep{BinoisGramacy2021}. We emphasize that HetGP is primarily designed for stochastic simulation settings with replicated design sites, where repeated observations directly inform local variance estimation, whereas our primary focus is large unreplicated spatial or temporal datasets. Variance smoothing in HetNV was performed in two dimensions using thin plate spline GAMs via the \texttt{mgcv} package \citep{wood2017}, and in one dimension using LOESS smoothing via the \texttt{loess} function in R \citep{Cleveland1988}.

Unless otherwise stated, all Vecchia-based models rescale inputs to the unit cube prior to neighborhood selection and use neighborhood size $m=20$. This choice reflects common practice in scalable Vecchia implementations, where neighborhood sizes between 20 and 30 typically provide a favorable tradeoff between likelihood approximation accuracy and computational cost \citep{guinness18,katzfuss21,katzfuss22}. In preliminary experiments, increasing $m$ beyond 20 yielded negligible improvements in predictive performance relative to the additional computational expense.

\paragraph{Length-scale and noise-amplitude ranges:}
Simulation parameters governing smoothness and noise magnitude are specified
as part of the data-generating process to span a range of practically relevant
signal-to-noise regimes.
Latent mean and noise variance functions are generated using smooth stochastic
processes with tunable regularity, and noise amplitudes are constrained to lie
within prescribed bounds $[e_{\min}, e_{\max}]$.
These settings induce a variety of effective length-scales and heteroskedastic
patterns without relying on model fitting or adaptive tuning, and are held
fixed across methods within each scenario.

\paragraph{Synthetic data generation:}
For computational efficiency in generating synthetic data, latent mean functions are generated on dense
lattices.
In two dimensions we employ a separable construction that permits sampling via
Kronecker eigendecompositions. This structure is used only for data generation;
all inference methods operate on unstructured input locations and do not assume
any grid or separability.

\paragraph{Metrics:}
RMSE of the mean is computed on a dense test grid against the latent mean function.
Empirical coverage is computed at the observation level using nominal 95\%
predictive intervals derived from the calibrated predictive distribution.
Let $f(\cdot)$ denote the latent mean function and let $\hat{\mu}(x)$ denote a model's posterior mean for $f(x)$.
On an evaluation set $\{x_j\}_{j=1}^N$ (dense grid in 1D; lattice in 2D), we report
\[
\mathrm{RMSE(mean)} \;=\; \sqrt{\frac{1}{N}\sum_{j=1}^N \big(\hat{\mu}(x_j)-f(x_j)\big)^2 } .
\]
For uncertainty calibration we compute observation-level coverage using the total predictive variance (defined below).
When the truth includes a standard-deviation surface, we also report variance-surface recovery metrics based on the
\emph{total predictive standard deviation} (not only the nugget); see below.

Let $\mathrm{tv}(x)$ denote the total predictive variance for an observation at input $x$,
i.e., kriging variance plus the (possibly input-dependent) nugget variance.
We define the total predictive standard deviation as
$s_{\mathrm{tot}}(x) = \sqrt{\mathrm{tv}(x)}$.
When the data-generating process provides a corresponding ground-truth standard deviation surface
$s_{\mathrm{tot,true}}(x)$ on the same evaluation set, we summarize recovery via
\[
\mathrm{RMSE(sd)} \;=\; \sqrt{\frac{1}{N}\sum_{j=1}^N \big(s_{\mathrm{tot}}(x_j)-s_{\mathrm{tot,true}}(x_j)\big)^2},
\qquad
\mathrm{Corr(sd)} \;=\; \mathrm{Corr}\big(s_{\mathrm{tot}}(x_j), s_{\mathrm{tot,true}}(x_j)\big).
\]
For homoskedastic baselines, $s_{\mathrm{tot}}(x)$ may vary through the kriging variance even if the nugget is constant.

Runtimes include hyperparameter optimization and variance estimation. The simulation protocol used for each configuration is summarized in Algorithm~\ref{alg:sim_protocol}.

\begin{algorithm}[H]
\caption{Simulation protocol per scenario (1D/2D; heteroskedastic)}
\label{alg:sim_protocol}
\Input{
Kernel smoothness set $\mathcal{V}=\{1.5,2.5,\infty\}$;
problem sizes $\mathcal{N}$;
replicate counts $R=100$;
noise-amplitude bounds $[e_{\min},e_{\max}]$.
}
\Output{
Mean $\pm$ sd of RMSE, empirical 95\% coverage, and runtime for each method.
}
\For{$\nu \in \mathcal{V}$}{
  \For{$n \in \mathcal{N}$}{
    \For{$r = 1$ \KwTo $R$}{
      Generate inputs and latent mean with a Matérn-$\nu$ kernel;\;
      Add heteroskedastic noise with amplitude in $[e_{\min},e_{\max}]$;\;
      Fit HomoSV and HetNV;
      fit HetGP  only if computationally feasible;\; %and deepGP
      Evaluate metrics on a held-out grid and record runtime.\;
    }
  }
}
Aggregate metrics over replicates and report mean $\pm$ standard deviation.

\end{algorithm}

Let $\mu_i$ denote the predictive mean for $y_i$ and let $\mathrm{tv}_i$ denote the corresponding
\emph{total predictive variance} for $y_i$ (kriging variance plus the nugget variance).
We compute standardized residuals $z_i = (y_i-\mu_i)/\sqrt{\mathrm{tv}_i}$ and report
empirical $\alpha$-coverage as
\[
\widehat{\mathrm{Cov}}(\alpha)=\frac{1}{n}\sum_{i=1}^n \mathbf{1}\!\left\{|z_i|\le z_{\alpha}\right\},
\qquad
z_{\alpha}=\Phi^{-1}\!\left(\frac{1+\alpha}{2}\right),
\]
with $\alpha=0.95$.
Coverage is interpreted as a calibration diagnostic rather than a quantity to maximize: values substantially above the nominal level indicate conservative intervals, while values below the nominal level indicate undercoverage. As additional calibration diagnostics we report $\frac{1}{n}\sum_i z_i^2$ and $\mathrm{median}(z_i^2)$, which equal 1 in expectation under a calibrated Gaussian predictive distribution.

\paragraph{Appendix materials }
Additional figures and diagnostics are provided in the Appendix to illustrate model behavior across simulation settings, including fitted surfaces, residual diagnostics, and replicate-level variability summaries.

\subsection{Simulation A: 1D Heteroskedastic}
\label{sec:sim-1d-hetero}

\paragraph{Setup:}
Inputs $x \in [0,1]$ are sampled on a regular grid.
The latent mean function $f$ is generated from a smooth stochastic process with
Matérn-$\nu$ regularity, and observations are corrupted by input-dependent noise
$\varepsilon \sim N(0,\sigma^2(x))$, where $\log \sigma^2(x)$ varies smoothly over
the input domain with amplitude constrained to $[e_{\min},e_{\max}]$.
We consider sample sizes $n \in \{100,\,1000,\,5000\}$.
We fit HomoSV and HetNV in all cases, and include HetGP only when computationally feasible. Computational feasibility was defined using a practical runtime threshold applied per model and configuration: for each combination of $(n,\nu)$ and model, we allowed up to 10 hours to complete all $R=100$ Monte Carlo replicates. Methods that exceeded this budget were excluded from that configuration.
All methods are evaluated on identical data realizations within each of the $R=100$ Monte Carlo replicates for each simulation setting.

\paragraph{Expected behavior:}
In this setting, the primary distinction among methods is their treatment of input-dependent noise. The homoskedastic baseline is expected to yield accurate point predictions for the mean function but misrepresent uncertainty when noise variance varies substantially across the input space. In contrast, HetNV adapts predictive uncertainty to local noise levels while preserving the scalability and local dependence structure of Vecchia approximations. HetGP serves as a reference heteroskedastic model at small sample sizes but becomes computationally prohibitive as $n$ grows. 

% The deepGP baseline offers flexible nonstationary modeling of the mean function, but assumes a homoskedastic noise variance and may incur substantially higher computational cost and optimization instability at larger sample sizes.

% \begin{figure}[htbp]
%   \centering
%   \includegraphics[width=\linewidth]{figures/Het1DTrue.png}
%  \caption{1D heteroskedastic simulation: A single MC replicate of a latent truth. 
% The solid black curve shows the true mean function $f(x)$, and the dashed black curves show $\pm 2\sigma(x)$, where $\sigma(x)$ is the true noise standard deviation. Points denote noisy observations.}
%   \label{fig:1d_truth}
% \end{figure}

% \begin{figure}[htbp]
%   \centering
%   \includegraphics[width=\linewidth]{figures/Het1DModel.png}
%   \caption{1D heteroskedastic simulation: model fits and predictive uncertainty for the simulated data realization from Figure \ref{fig:1d_truth}. 
% For each method we show the posterior mean $\hat f(x)$ and pointwise 95\% predictive intervals for $y$, combining kriging variance and observation-level noise variance. 
% HomoSV uses a constant noise variance across the domain, whereas HetNV adapts interval width through $\hat\sigma^2(x)$. 
% HetGP provides a latent-variance benchmark when feasible.}
% % , and deepGP reflects a flexible nonstationary alternative.}

%   \label{fig:1d_modelfits}
% \end{figure}

\begin{table}[H]
\centering
\begin{tabular}{|c|c|l|c|c|c|}
\hline
\textbf{$n$} & \textbf{$\nu$} & \textbf{Model} &
\textbf{RMSE(mean)} &
\textbf{Coverage} &
\textbf{Time (s)} \\
\hline

\multirow{9}{*}{\centering 100}
& \multirow{3}{*}{\centering 1.5}
  & HomoSV & 0.105 $\pm$ 0.004          & 0.896 $\pm$ 0.004 & \textbf{0.278} $\pm$ 0.054\\
& & HetNV  & 0.097 $\pm$ 0.004 & \textbf{0.898} $\pm$ 0.003 & 0.610 $\pm$ 0.088 \\
& & HetGP  & \textbf{0.096} $\pm$ 0.003          & 0.752 $\pm$ 0.006 & 0.345 $\pm$ 0.049\\
% & & deepGP & 0.072 $\pm$ 0.018          & 0.997 $\pm$ 0.014           & 38.903 $\pm$ 1.820 \\
\cline{2-6}
& \multirow{3}{*}{\centering 2.5}
  & HomoSV & 0.095 $\pm$ 0.005 & 0.896 $\pm$ 0.004  & 0.863 $\pm$ 0.109\\ 

& &  HetNV  & 0.090 $\pm$0.006 & \textbf{0.902} $\pm$0.003 & 0.598 $\pm$ 0.122 \\
& & HetGP  & \textbf{0.089} $\pm$ 0.004  & 0.745 $\pm$ 0.008    & \textbf{0.383} $\pm$ 0.057 \\
% & & deepGP & 0.072 $\pm$ 0.018          & 0.997 $\pm$ 0.014           & 40.659 $\pm$ 1.860 \\
\cline{2-6}
& \multirow{3}{*}{\centering $\infty$}
& HomoSV & 0.141 $\pm$0.019 & 0.875 $\pm$0.010 & 3.462 $\pm$ 0.620 \\
&  & HetNV  & 0.129 $\pm$ 0.019 & \textbf{0.896} $\pm$0.006 & 6.056 $\pm$ 0.833 \\
& & HetGP  & \textbf{0.090} $\pm$0.006 & 0.741  $\pm$ 0.009 & \textbf{0.248} $\pm$ 0.041   \\
% & & deepGP & 0.072 $\pm$ 0.018          & 0.997 $\pm$ 0.014           & 40.522 $\pm$ 1.895 \\
\hline

% ===================== n = 1000 =====================
\multirow{9}{*}{\centering 1000}
& \multirow{3}{*}{\centering
1.5}
& HomoSV & 0.041 $\pm$ 0.010 &  0.880 $\pm$ 0.003 & \textbf{10.613} $\pm$ 2.098 \\
& & HetNV  & \textbf{0.040} $\pm$ 0.008 & \textbf{0.903} $\pm$ 0.002 & 20.31 $\pm$ 1.99 \\
& & HetGP  & 0.062 $\pm$ 0.004      & 0.954 $\pm$ 0.021 & 286.802 $\pm$ 153.533 \\
% & & deepGP & 0.049 $\pm$ 0.012          & 1.000 $\pm$ 0.000           & 254.676 $\pm$ 3.649 \\
\cline{2-6}

& \multirow{3}{*}{\centering
 2.5}
& HomoSV & \textbf{0.039} $\pm$ 0.005 & 0.881 $\pm$ 0.002 & \textbf{0.986 $\pm$ 0.208} \\
& & HetNV  & \textbf{0.039} $\pm$ 0.004 & \textbf{0.904} $\pm$ 0.002 & 4.536 $\pm$ 0.520 \\
& & HetGP  & 0.057 $\pm$ 0.014 & 0.959 $\pm$ 0.025 & 305.078 $\pm$ 153.992 \\
% & & deepGP & 0.049 $\pm$ 0.012          & 1.000 $\pm$ 0.000           & 255.246 $\pm$ 4.656 \\
\cline{2-6}

& \multirow{3}{*}{\centering $\infty$}
  & HomoSV & 0.528 $\pm$ 0.1425 & 0.899 $\pm$ 0.002 & \textbf{2.264 }$\pm$ 0.030 \\
&  & HetNV  & \textbf{0.474} $\pm$ 0.182 & 0.894 $\pm$ 0.001 & 4.773 $\pm$ 0.049 \\
& & HetGP  & 0.501 $\pm$ 0.006  &\textbf{ 0.944} $\pm$ 0.031  & 336.139 $\pm$  203.554 \\
% & & deepGP & 0.049 $\pm$ 0.012          & 1.000 $\pm$ 0.000           & 259.151 $\pm$ 6.729 \\
\hline

% ===================== n = 5000 =====================
\multirow{9}{*}{\centering 5000}
& \multirow{3}{*}{\centering 1.5}
 & HomoSV & \textbf{0.020} $\pm$ 0.003 & 0.842 $\pm$ 0.047 & \textbf{26.866} $\pm$ 27.582 \\
& & HetNV  & 0.022 $\pm$ 0.003 & \textbf{0.865} $\pm$ 0.037 & 78.681 $\pm$ 30.136 \\
& & HetGP  & -- & -- & -- \\
% & & deepGP & -- & -- & -- \\
\cline{2-6}

& \multirow{3}{*}{\centering 2.5}
& HomoSV & 0.132 $\pm$ 0.053 & 0.726 $\pm$ 0.104 & \textbf{2.081} $\pm$ 0.227 \\
&  & HetNV  & \textbf{0.112} $\pm$ 0.006 & \textbf{0.897} $\pm$ 0.095 & 32.785 $\pm$ 0.783 \\
& & HetGP  & -- & -- & -- \\
% & & deepGP & -- & -- & -- \\
\cline{2-6}

& \multirow{3}{*}{\centering $\infty$}
& HomoSV & 0.278 $\pm$ 0.066 & 0.840 $\pm$ 0.085 & \textbf{8.402} $\pm$ 1.019 \\
& & HetNV  & \textbf{0.258} $\pm$ 0.057 & \textbf{0.937} $\pm$ 0.134 & 28.042 $\pm$ 0.503 \\
& & HetGP  & -- & -- & -- \\
% & & deepGP & -- & -- & -- \\
\hline

\end{tabular}
\caption{1D heteroskedastic setting. RMSE(mean), observation-level coverage, and computation time (mean $\pm$ MC standard deviation). Best RMSE per $(n,\nu)$ block is bolded. Differences across $\nu$ occur at higher precision but are not practically meaningful at the reported scale. HetGP was not run for the larger $n=5000$ case because of the computational expense.}

\label{table:1d_hetero_mean}
\end{table}

\begin{table}[H]
\centering
\begin{tabular}{|c|c|l|c|c|}
\hline
\textbf{$n$} & \textbf{$\nu$} & \textbf{Model} &
\textbf{RMSE($s_{\mathrm{tot}}$)} &
\textbf{Corr($s_{\mathrm{tot}}$)} \\
\hline

% ===================== n = 100 =====================
% \multirow[c]{9}{*}{100}x
\multirow{9}{*}{\centering 100}

& \multirow{3}{*}{\centering 1.5}
  & HetNV  & \textbf{0.131} $\pm$ 0.004 & \textbf{0.806} $\pm$ 0.015 \\
& & HomoSV &0.155 $\pm$0.003 & -0.092 $\pm$ 0.036  \\
& & HetGP  & 0.187 $\pm$ 0.004          & 0.745 $\pm$ 0.033 \\
% & & deepGP & 0.124 $\pm$ 0.009          & 0.052 $\pm$ 0.282 \\
\cline{2-5}

& \multirow[c]{3}{*}{\centering 2.5}
      & HetNV  & \textbf{0.12} $\pm$0.004 & \textbf{ 0.816} $\pm$0.015 \\
& & HomoSV & 0.151 $\pm$0.002 &  -0.084 $\pm$0.046 \\
& & HetGP  & 0.199 $\pm$0.006 &  0.628 $\pm$ 0.055 \\
% & & deepGP & 0.124 $\pm$ 0.009          & 0.052 $\pm$ 0.282 \\
\cline{2-5}

& \multirow[c]{3}{*}{\centering $\infty$}
  & HetNV  & \textbf{0.116} $\pm$ 0.003 &   \textbf{0.774} $\pm$ 0.020 \\
& & HomoSV & 0.151 $\pm$ 0.003 &   -0.105 $\pm$ 0.037 \\
& & HetGP  & 0.197 $\pm$ 0.006 &   0.728 $\pm$ 0.045 \\
% & & deepGP & 0.124 $\pm$ 0.009          & 0.052 $\pm$ 0.282 \\
\hline
% ===================== n = 1000 =====================
\multirow{9}{*}{\centering 1000}
& \multirow{3}{*}{\centering 1.5}
  & HetNV  &  \textbf{0.117} $\pm$ 0.001        &  0.902  $\pm$ 0.004 \\
& & HomoSV & 0.156 $\pm$ 0.002          &  -0.118 $\pm$ 0.023 \\
& & HetGP  &  0.187 $\pm$ 0.002 & \textbf{0.930} $\pm$ 0.014 \\
% & & deepGP & 0.127 $\pm$ 0.003          & 0.023 $\pm$ 0.372 \\
\cline{2-5}

& \multirow{3}{*}{\centering 2.5}
  & HetNV  & \textbf{0.123} $\pm$ 0.002        & \textbf{0.959} $\pm$ 0.009 \\
& & HomoSV & 0.156 $\pm$ 0.001          &   -0.008  $\pm$ 0.014 \\
& & HetGP  &  0.183 $\pm$ 0.003 & 0.920 $\pm$ 0.009 \\
% & & deepGP & 0.127 $\pm$ 0.003          & 0.023 $\pm$ 0.372 \\
\cline{2-5}

& \multirow{3}{*}{\centering $\infty$}
  & HetNV  & \textbf{0.156} $\pm$ 0.001          &  \textbf{0.937} $\pm$ 0.007 \\
& & HomoSV & 0.177 $\pm$ 0.002         &  -0.018 $\pm$ 0.010 \\
& & HetGP  & 0.183 $\pm$ 0.005 &  \textbf{0.937} $\pm$ 0.008\\
% & & deepGP & 0.127 $\pm$ 0.003          & 0.023 $\pm$ 0.372 \\
\hline

% ===================== n = 5000 =====================
\multirow{9}{*}{\centering 5000}
& \multirow{3}{*}{\centering 1.5}
  & HetNV  & \textbf{0.120} $\pm$ 0.002 & \textbf{0.767 }$\pm$ 0.015 \\
& & HomoSV & 0.155$\pm$ 0.002          &  -0.051 $\pm$ 0.045 \\
& & HetGP  & -- & --\\

\cline{2-5}

& \multirow{3}{*}{\centering 2.5}
  & HetNV  & \textbf{0.146} $\pm$ 0.016 & \textbf{0.755} $\pm$ 0.010 \\
& & HomoSV & 0.152 $\pm$ 0.002       &  0.002 $\pm$ 0.018 \\
& & HetGP  & -- & -- \\

\cline{2-5}

& \multirow{3}{*}{\centering $\infty$}
  & HetNV  & \textbf{0.315} $\pm$ 0.154 & \textbf{0.542} $\pm$ 0.147 \\
& & HomoSV & 0.319 $\pm$ 0.139          &  -0.001 $\pm$ 0.010 \\
& & HetGP  & -- & -- \\

\hline

\end{tabular}

\caption{1D heteroskedastic setting. RMSE and correlation for the total predictive standard deviation surface $s_{\mathrm{tot}}(x)=\sqrt{\mathrm{tv}(x)}$ (mean $\pm$ MC sd). Bold entries match the best values per $(n,\nu)$ block. HetGP was not run for larger n case because of the computational expense. }

% HetGP and deepGP were not run for the larger $n=5000$ case because of their computational expense.}

\label{table:1d_hetero_sd}
\end{table}

\begin{figure}[H]
   \centering
   \makebox[\linewidth][c]{%
     \includegraphics[trim=0.05cm 0.05cm 0.05cm 0.05cm,clip=true,width=1.2\linewidth]{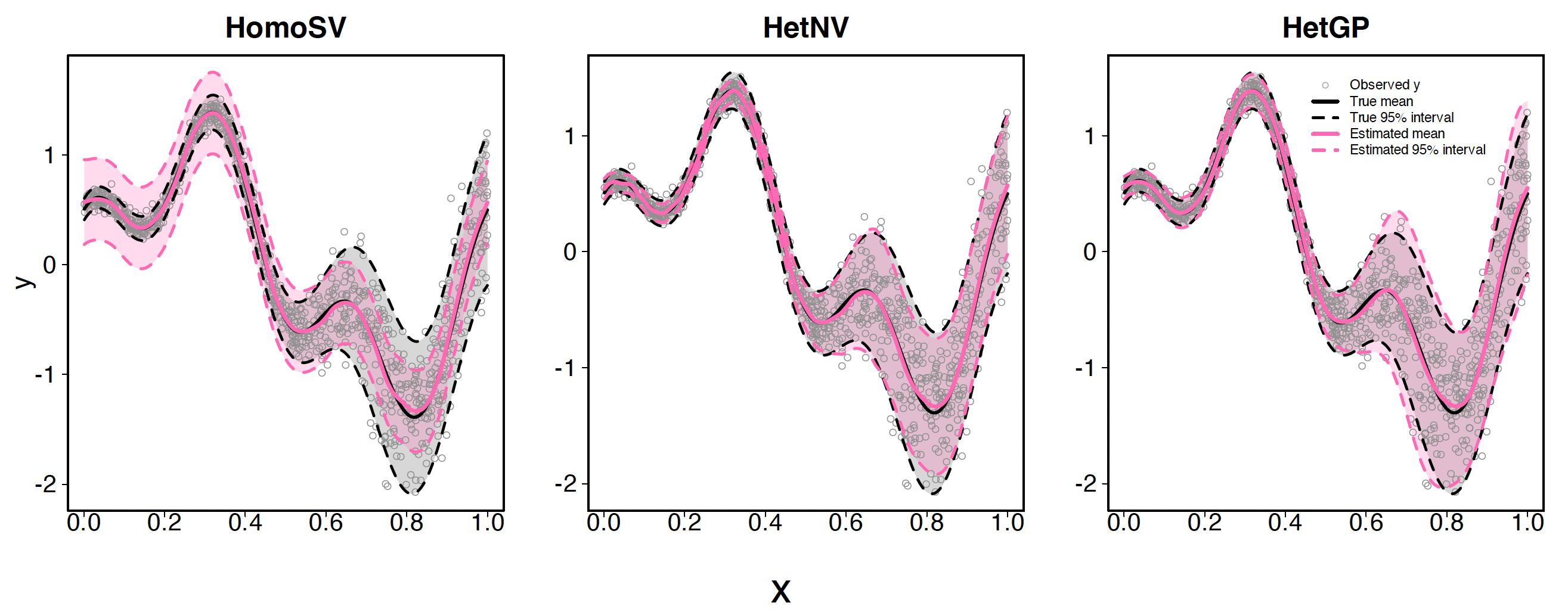}
   }
   \caption{One-dimensional heteroskedastic example illustrating model behavior for a sample size of $n=1000$. The solid black curve denotes the true latent mean function, and the dashed black curves show the corresponding 95\% observation-noise intervals. Points represent noisy observations. Colored curves show the estimated posterior mean and corresponding 95\% plug-in predictive intervals for each model, with predictive intervals shown as dashed lines. HetNV adapts interval width to local noise levels, narrowing in low-noise regions and widening in high-noise regions, while HomoSV produces more spatially uniform uncertainty. HetGP provides a flexible heteroskedastic benchmark at this sample size.}
   \label{fig:1d_truth}
\end{figure}

\paragraph{Results and interpretation:}

Table~\ref{table:1d_hetero_mean} summarizes mean-function accuracy, empirical coverage, and runtime. HetNV achieves comparable RMSE for the latent mean function relative to HomoSV across all configurations, indicating that modeling heteroskedasticity does not degrade mean estimation. However, HetNV substantially improves uncertainty calibration by adapting predictive interval widths to local noise levels, whereas HomoSV relies on a single global variance.

Table~\ref{table:1d_hetero_sd} focuses on recovery of the total predictive standard deviation surface. 
Here, HetNV substantially improves correlation with the true variability surface and reduces RMSE relative to the homoskedastic baseline. 
These results confirm that the primary advantage of HetNV lies in improved uncertainty quantification rather than changes in mean-surface reconstruction.

At $n=5000$ both HomoSV and HetNV under-cover, which is consistent with approximation and/or optimization error dominating uncertainty quantification at this fixed neighborhood size $m$.
In particular, when $m$ is held constant as $n$ increases, Vecchia approximation error can accumulate, and coverage can degrade even when point RMSE remains stable.

Noise-surface diagnostics in Table~\ref{table:1d_hetero_sd} show that HetNV more accurately recovers spatial variation in $\sigma(x)$, with lower RMSE in $\hat\sigma(x)$ and substantially higher correlation with the true total predictive SD surface process compared to the homoskedastic baseline. Runtime results show that HetNV retains the near-linear scaling of Vecchia inference (for fixed neighborhood size $m$), with modest additional overhead from variance smoothing.

% Figures~\ref{fig:1d_truth} and \ref{fig:1d_modelfits} illustrate these effects visually. 
% Figure~\ref{fig:1d_truth} shows the latent mean and true variability structure, highlighting regions of low and high noise. 
As shown in Figure~\ref{fig:1d_truth}, HomoSV produces predictive intervals of nearly constant width, leading to over-dispersion in low-noise regions and under-dispersion in high-noise regions. 
In contrast, HetNV widens intervals in high-noise regions and narrows them in low-noise regions, yielding predictive bands that are neither uniformly over- nor under-dispersed while preserving a smooth mean estimate. 
When feasible, HetGP exhibits similar qualitative adaptation at small $n$ but at substantially greater computational cost.

Overall, these results show that the primary benefit of HetNV in one dimension is improved uncertainty quantification, with limited impact on mean prediction accuracy.

\subsection{Simulation B: 2D Heteroskedastic}
\label{sec:sim-2d-hetero}

\paragraph{Setup:}
Inputs $x = (x_1,x_2) \in [0,1]^2$ are sampled on regular two-dimensional lattices with grid resolutions $25\times25$, $50\times50$, and $100\times100$. The latent mean function $f$ is generated from a smooth stochastic process with Matérn-$\nu$ regularity. For computational efficiency in data generation only, a separable construction is used; inference methods do not assume separability or grid structure. Observations are corrupted by input-dependent noise $\varepsilon \sim N(0,\sigma^2(x))$, where $\log \sigma^2(x)$ varies smoothly over the input domain with amplitude constrained to $[e_{\min},e_{\max}]$.

\begin{figure}[htbp]
  \centering
  % \begin{subfigure}[t]{0.68\textwidth}
    \centering
    \includegraphics[width=\linewidth]{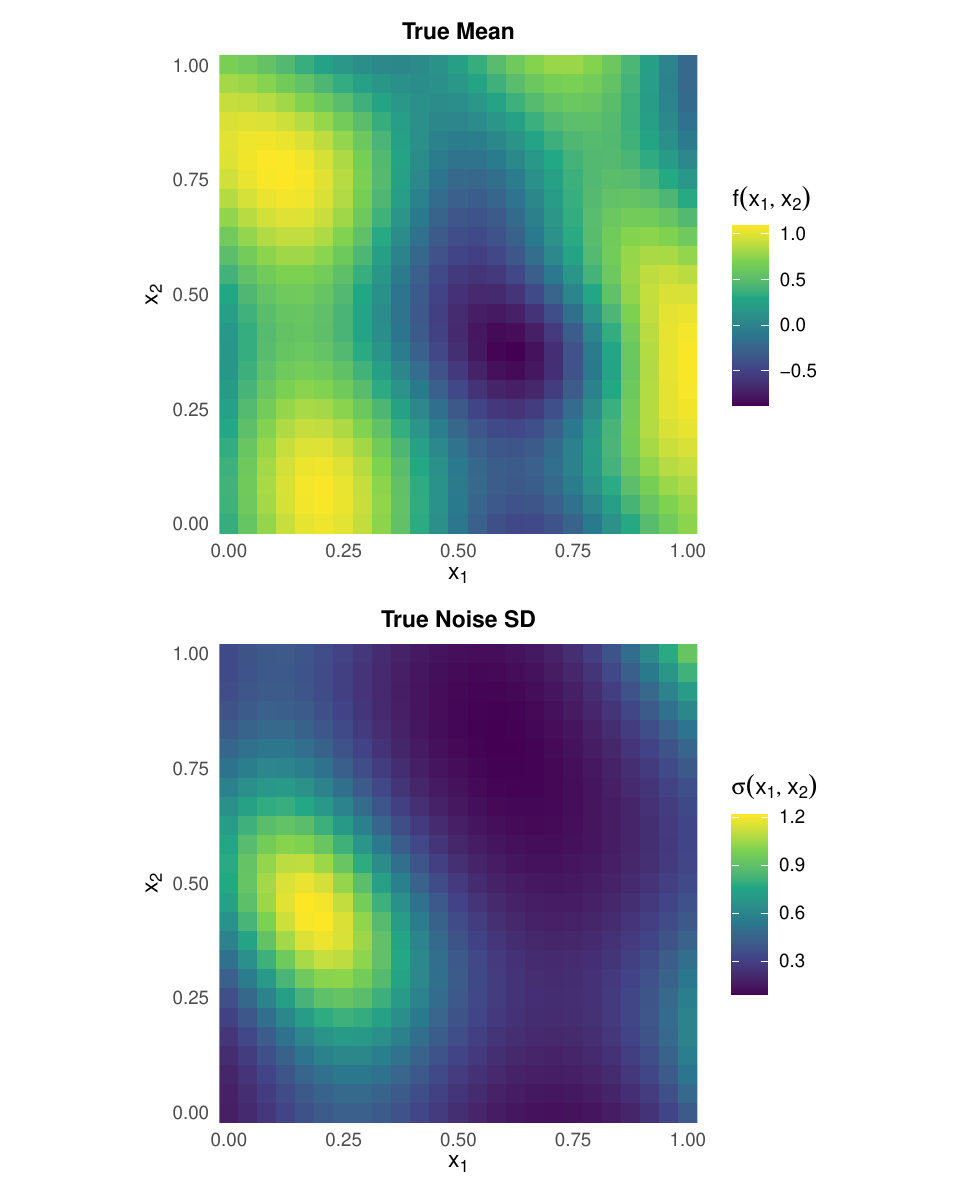}
    \label{fig:truemean}
  % \end{subfigure}\hfill
  % \begin{subfigure}[t]{0.68\textwidth}
  %   \centering
  %   \includegraphics[width=\linewidth]{figures/TrueNoiseSD2d.png}
  % \caption{True Noise SD}
  %   \label{fig:truenoise}  
  %   \end{subfigure}
  \caption{Representative realization from the two-dimensional heteroskedastic simulation. The top panel shows the latent mean surface, and the bottom panel shows the input-dependent noise standard deviation. The example illustrates the setting in which a single global nugget cannot represent the spatially varying observation noise.}\label{fig:truemeanandsd}
\end{figure}

We fit HomoSV and HetNV at all grid resolutions. A more computationally intensive baseline (HetGP) is included only at smaller grid sizes where feasible. All methods are evaluated on identical data realizations within each Monte Carlo replicate; we perform $R=50$ total replicates per setting for the smaller grids, and $R=25$ for the largest grid. An example realization of a true mean and noise standard deviation surface is shown in Figure~\ref{fig:truemeanandsd}. Coverage is computed at observed locations using total predictive variance (kriging variance plus nugget), via standardized residuals as defined above.

\paragraph{Expected behavior:}
This simulation tests whether scalable GP approximations can represent strongly spatially varying uncertainty in two dimensions.
When heteroskedasticity is mild, mean RMSE may be similar across methods.
However, when noise variation is strong, a homoskedastic likelihood can mis-weight observations across the domain, distort hyperparameter estimates, and consequently degrade both uncertainty calibration and mean-surface reconstruction.
HetNV is designed to mitigate this by learning an input-dependent nugget surface and reweighting the mean fit accordingly.
\begin{table}[H]
\centering
\begin{tabular}{|c|c|l|c|c|c|}
\hline
\textbf{$n_x \times n_y$} & \textbf{$\nu$} & \textbf{Model} &
\textbf{RMSE(mean)} & \textbf{Coverage} & \textbf{Time (s)} \\
\hline

\multirow{9}{*}{\centering25x25}
& \multirow{3}{*}{1.5}
& HomoSV & 0.211 $\pm$ 0.048 & 0.928 $\pm$ 0.021 & \textbf{2.373} $\pm$ 0.660 \\
& & HetNV   & 0.177 $\pm$ 0.045 & \textbf{0.931} $\pm$ 0.016 & 15.326 $\pm$ 3.628 \\
& & HetGP & \textbf{0.151} $\pm$ 0.041 & 0.869 $\pm$ 0.030 & 29.646 $\pm$ 17.179 \\
\cline{2-6}

& \multirow{3}{*}{\centering2.5}
& HomoSV & 0.151 $\pm$ 0.031 & \textbf{0.959} $\pm$ 0.014 & \textbf{2.534} $\pm$ 0.950 \\
& & HetNV  & 0.118 $\pm$ 0.023 & 0.919 $\pm$ 0.018 & 15.243 $\pm$ 4.146 \\
& & HetGP  & \textbf{0.095} $\pm$ 0.025 & 0.746 $\pm$ 0.050 & 26.517 $\pm$ 15.342 \\
\cline{2-6}

& \multirow{3}{*}{\centering$\infty$}
& HomoSV& 0.144 $\pm$ 0.112 & 0.790 $\pm$ 0.018 & \textbf{2.600} $\pm$ 0.550 \\
& & HetNV  & 0.071 $\pm$ 0.024 & \textbf{0.872} $\pm$ 0.024 & 15.051 $\pm$ 10.709 \\
& & HetGP  & \textbf{0.056} $\pm$ 0.017 & 0.614 $\pm$ 0.054 & 24.618 $\pm$ 12.276 \\
\cline{2-6}
\hline

\multirow{9}{*}{\centering50x50}
& \multirow{2}{*}{1.5}
& HomoSV & 0.158 $\pm$ 0.029 & \textbf{0.952} $\pm$ 0.015 & \textbf{5.350} $\pm$ 1.298 \\
& & HetNV  & 0.128 $\pm$ 0.027 & 0.927 $\pm$ 0.012 & 33.201 $\pm$ 4.427 \\
& & HetGP  & \textbf{0.092} $\pm$ 0.018 & 0.837 $\pm$ 0.024 & 113.625 $\pm$ 80.898 \\
\cline{2-6}

& \multirow{3}{*}{\centering2.5}
& HomoSV & 0.105 $\pm$ 0.020 & \textbf{0.976} $\pm$ 0.009 & \textbf{5.000} $\pm$ 1.308 \\
& & HetNV  & 0.085 $\pm$ 0.018 & 0.909 $\pm$ 0.015 & 31.036 $\pm$ 4.629 \\
& & HetGP  & \textbf{0.059} $\pm$ 0.013 & 0.680 $\pm$ 0.052 & 164.364 $\pm$ 92.752 \\
\cline{2-6}

& \multirow{3}{*}{\centering$\infty$}
& HomoSV& 0.112 $\pm$ 0.096 & \textbf{0.993} $\pm$ 0.012 & \textbf{11.352} $\pm$ 2.505 \\
& & HetNV  & 0.045 $\pm$ 0.013 & 0.858 $\pm$ 0.018 & 32.089 $\pm$ 4.851 \\
& & HetGP  &  \textbf{0.034} $\pm$ 0.008 & 0.519 $\pm$ 0.049 & 161.990 $\pm$ 87.329 \\
\cline{2-6}
\hline

\multirow{9}{*}{\centering 100x100}
& \multirow{2}{*}{\centering1.5}
& HomoSV & 0.124 $\pm$ 0.021 & \textbf{0.968} $\pm$ 0.010 & \textbf{14.771} $\pm$ 3.550 \\
& & HetNV  & \textbf{0.096} $\pm$ 0.019 & 0.925 $\pm$ 0.009 & 85.520 $\pm$ 18.807 \\
& & HetGP  & -- & -- & -- \\
\cline{2-6}

& \multirow{3}{*}{\centering2.5}
& HomoSV & 0.078 $\pm$ 0.014 & \textbf{0.983} $\pm$ 0.005 & \textbf{20.004} $\pm$ 8.652 \\
& & HetNV  & \textbf{0.060} $\pm$ 0.009 & 0.958 $\pm$ 0.013 & 93.965 $\pm$ 22.606 \\
& & HetGP  & -- & -- & -- \\
\cline{2-6}

& \multirow{3}{*}{\centering$\infty$}
 & HomoSV& 0.103 $\pm$ 0.066 & 0.998 $\pm$ 0.012 & \textbf{39.584} $\pm$ 16.815 \\
&& HetNV   & \textbf{0.072} $\pm$ 0.011 & \textbf{0.993} $\pm$ 0.009 & 108.430 $\pm$ 33.818 \\
& & HetGP  & -- & -- & -- \\
\cline{2-6}
\hline

\end{tabular}

\caption{2D heteroskedastic setting. RMSE(mean), observation-level 95\% coverage, and computation time (mean $\pm$ MC standard deviation). Bold RMSE entries indicate the lowest mean RMSE within each $(n_x \times n_y,\nu)$ block. Coverage should be interpreted relative to the nominal 95\% level rather than as a quantity to maximize. HetGP was omitted for the $100\times100$ grid because of computational expense.}
\label{table:2d_hetero_mean}
\end{table}

\begin{table}[H]
\centering
\begin{tabular}{|c|c|l|c|c|}
\hline
\textbf{$n_x \times n_y$} & \textbf{$\nu$} & \textbf{Model} &
\textbf{RMSE($s_{\mathrm{tot}}$)} & \textbf{Corr($s_{\mathrm{tot}}$)} \\
\hline

% ===================== 25x25 =====================
\multirow{9}{*}{\centering 25x25}
& \multirow{3}{*}{\centering 1.5}
& HomoSV & 0.324 $\pm$ 0.018 &  -0.003 $\pm$ 0.048\\
& & HetNV  & \textbf{0.147} $\pm$ 0.009 & \textbf{0.476} $\pm$ 0.113 \\
& & HetGP  & 0.196 $\pm$ 0.042 & 0.250 $\pm$ 0.065 \\
\cline{2-5}

& \multirow{3}{*}{\centering 2.5}
& HomoSV & 0.325 $\pm$ 0.021 &  0.007 $\pm$ 0.044\\ 
& & HetNV  & \textbf{0.146} $\pm$ 0.003 & \textbf{0.631} $\pm$ 0.062 \\
& & HetGP  & 0.149 $\pm$ 0.017 & 0.536 $\pm$ 0.123 \\
% & & deepGP & 0.340 $\pm$ 0.830 & 0.312 $\pm$ 0.160 \\
\cline{2-5}

& \multirow{3}{*}{\centering $\infty$}
& HomoSV & 0.440 $\pm$ 0.376 &  -0.004 $\pm$ 0.051\\ 
& & HetNV  & \textbf{0.147} $\pm$ 0.002 & \textbf{0.687} $\pm$ 0.051 \\
& & HetGP  & 0.149 $\pm$ 0.010 & 0.534 $\pm$ 0.139 \\
% & & deepGP & 0.331 $\pm$ 0.828 & 0.283 $\pm$ 0.171 \\
\hline

% ===================== 50x50 =====================
\multirow{9}{*}{\centering 50x50}
& \multirow{3}{*}{1.5}
& HomoSV & 0.324 $\pm$ 0.021 &  0.010 $\pm$ 0.055\\ 
& & HetNV  & \textbf{0.131} $\pm$ 0.003 & \textbf{0.661} $\pm$ 0.083 \\
& & HetGP  & 0.140 $\pm$ 0.011 & 0.337 $\pm$ 0.075 \\

\cline{2-5}

& \multirow{3}{*}{\centering 2.5}
& HomoSV & 0.321 $\pm$ 0.023 &  -0.006 $\pm$ 0.050\\ 
& & HetNV  & \textbf{0.131} $\pm$ 0.003 & \textbf{0.761} $\pm$ 0.035 \\
& & HetGP  & 0.132 $\pm$ 0.007 & 0.651 $\pm$ 0.166 \\
\cline{2-5}

& \multirow{3}{*}{\centering $\infty$}
& HomoSV & 0.321 $\pm$ 0.023 & 0.001 $\pm$ 0.061\\
& & HetNV  & \textbf{0.133} $\pm$ 0.003 & \textbf{0.786} $\pm$ 0.026 \\
& & HetGP  & 0.135 $\pm$ 0.004 & 0.604 $\pm$ 0.150 \\

\hline

% ===================== 100x100 =====================
\multirow{9}{*}{\centering 100x100}
& \multirow{3}{*}{\centering 1.5}
& HomoSV & 0.324 $\pm$ 0.019 & 0.001 $\pm$ 0.056 \\
& & HetNV  & \textbf{0.122} $\pm$ 0.002 & \textbf{0.770} $\pm$ 0.055 \\
& & HetGP & -- & -- \\
\cline{2-5}

& \multirow{3}{*}{\centering 2.5}
& HomoSV & 0.324 $\pm$ 0.021 & -0.003 $\pm$ 0.055 \\
& & HetNV  & \textbf{0.123} $\pm$ 0.002 & \textbf{0.832} $\pm$ 0.021 \\
& & HetGP & -- & -- \\

\cline{2-5}

& \multirow{3}{*}{\centering $\infty$}
& HomoSV & 0.323 $\pm$ 0.025 & 0.001 $\pm$ 0.057 \\
& & HetNV  & \textbf{0.123} $\pm$ 0.002 & \textbf{0.844} $\pm$ 0.018 \\
& & HetGP & -- & -- \\

\hline

\end{tabular}

\caption{2D heteroskedastic setting. RMSE and correlation for the total predictive standard deviation surface $s_{\mathrm{tot}}(x)=\sqrt{\mathrm{tv}(x)}$ (mean $\pm$ MC sd). 
HetNV consistently achieves substantially higher correlation with the true variability surface than HomoSV across grid sizes. Best RMSE values per $(n_x \times n_y,\nu)$ block are bolded. Coverage values should be interpreted relative to the nominal 95\% level. Differences across $\nu$ occur at higher precision but are not practically meaningful at the reported scale.}
\label{table:2d_hetero_sd}
\end{table}

\begin{figure}[htbp]
  \centering
  % \begin{subfigure}[t]{0.68\textwidth}
    % \centering
    \includegraphics[width=\linewidth]{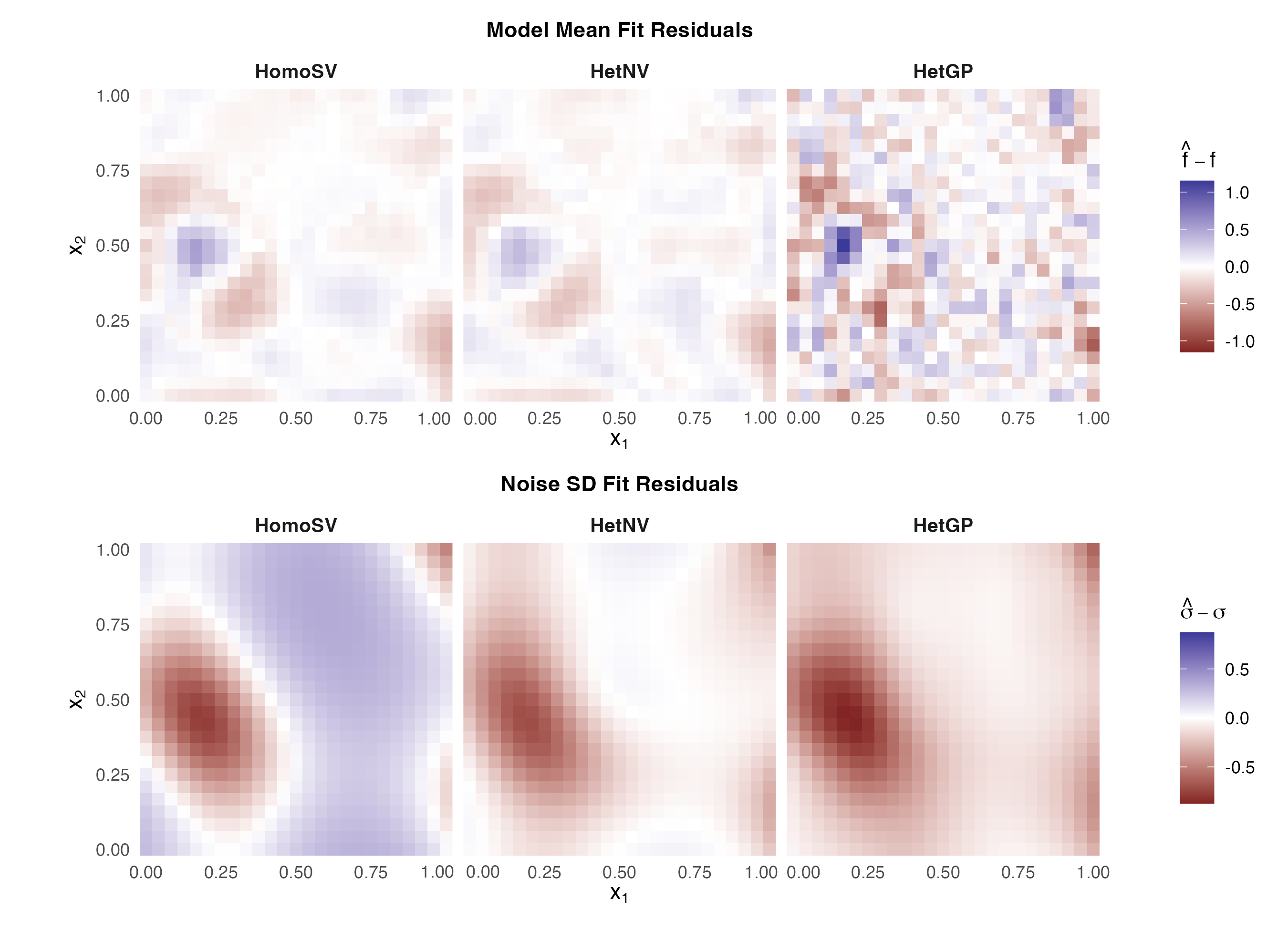}
    % \caption{Model Mean Fit Residuals and Noise SD Fit Residuals}
    
  % \end{subfigure}\hfill
  % \begin{subfigure}[t]{0.68\textwidth}
    % \centering
    % \includegraphics[width=\linewidth]{figures/noisesdresids.png}
  % \caption{}
    % % \label{fig:noisesdresids}  
    % \end{subfigure}
    \caption{Model-fit residual diagnostics for the representative two-dimensional simulation in Figure \ref{fig:truemeanandsd}. The top row shows latent mean residuals \(\hat f(x_1,x_2)-f(x_1,x_2)\), and the bottom row shows noise standard deviation residuals \(\hat\sigma(x_1,x_2)-\sigma(x_1,x_2)\). HomoSV cannot represent spatially varying noise and therefore leaves structured residual error in the variance surface. HetNV reduces this structure while maintaining a smoother mean-residual field than the dense heteroskedastic GP fit in this replicate.}
    \label{fig:meanfitresids}
\end{figure}

% Additional model-fit visualizations and diagnostics for this scenario are provided in Appendix~\ref{sec:appendix:additional_figs}.
\paragraph{Results and interpretation:}

Tables~\ref{table:2d_hetero_mean} and~\ref{table:2d_hetero_sd} show that heteroskedastic modeling has a substantial impact in two-dimensional settings. HetNV reduces mean RMSE relative to HomoSV across all grid resolutions and smoothness settings, with the largest relative improvements occurring for smoother latent surfaces. This suggests that the input-dependent nugget affects more than interval width: by downweighting high-noise regions, the heteroskedastic fit changes covariance estimation and improves recovery of the latent mean surface. Thus, in two dimensions, HetNV improves both uncertainty representation and mean-function reconstruction.

For example, on the \(100\times100\) grid, HetNV reduces RMSE(mean) relative to HomoSV from 0.124 to 0.096 for \(\nu=1.5\), from 0.078 to 0.060 for \(\nu=2.5\), and from 0.103 to 0.072 for \(\nu=\infty\). These correspond to relative reductions of approximately 23\%, 23\%, and 30\%, respectively. 

% \emph{Coverage and calibration.}
Coverage results should be interpreted carefully. HomoSV frequently attains coverage at or above the nominal 95\% level, but this often reflects conservative overcoverage from a single global nugget rather than locally accurate uncertainty. HetNV generally produces narrower, spatially adaptive intervals. In some settings, especially for $\nu=2.5$ and $\nu=\infty$, this leads to undercoverage. This is consistent with the plug-in nature of the method: uncertainty in the estimated variance surface is not fully propagated into the final predictive distribution. Thus, coverage alone does not fully summarize the benefit of HetNV; it must be considered together with recovery of the variance structure. This undercoverage is expected in plug-in approaches where uncertainty in the estimated variance surface is not propagated.

% \emph{Noise-surface recovery.}
Table~\ref{table:2d_hetero_sd} shows the clearest advantage of HetNV. Across all grid resolutions and smoothness settings, HetNV substantially reduces RMSE for the estimated noise standard deviation surface and yields positive, often strong, correlation with the true heteroskedastic structure. The correlations generally increase with grid resolution, indicating that the residual-based GAM smoother benefits from denser spatial information. In contrast, HomoSV cannot represent 
spatially varying noise, so its correlation with the heteroskedastic truth is not meaningful.

% \emph{Comparison with HetGP.}
Where feasible, HetGP serves as a useful small-scale heteroskedastic benchmark, although its underlying replication-oriented formulation differs from the large unreplicated setting targeted by HetNV. It sometimes achieves lower mean RMSE than HetNV, particularly for smoother surfaces on the smaller grids, but it often exhibits substantial undercoverage and much higher computational cost. Since HetGP was not feasible for the largest grid, we treat it as a small-scale benchmark rather than a scalable competitor.

% \emph{Computation.}
HetNV is slower than HomoSV because it adds a nonparametric variance-smoothing step, but it remains feasible for the largest grid considered. For fixed neighborhood size $m$, the dominant cost remains the Vecchia mean update, with additional overhead from the GAM variance fit.

% \emph{Visual diagnostics.}
Figures~\ref{fig:meanfitresids} illustrate the same pattern visually. HetNV reduces structured error in the estimated noise surface and produces mean residuals with less spatial patterning than the homoskedastic fit, consistent with improved local weighting under heteroskedasticity.

These results highlight that the benefits of heteroskedastic modeling extend beyond uncertainty calibration and can materially improve predictive performance in higher-dimensional settings.

\section{Case Study: In-Situ Plasma Readings}
\label{sec:aps}

We illustrate HetNV on time--energy spectrograms from the \emph{Autonomous Plume Sentry} (APS), an in-situ plasma diagnostic that records current as a function of time and particle energy in low-SNR environments \citep{ulrich2026epee,Maldonado2023APS}. APS provides a demanding testbed for heteroskedastic modeling: (i) measured currents can approach instrument sensitivity, (ii) spectral shapes vary with operating context, and (iii) the cadence and volume preclude exact GP inference. 

The application is computationally representative of the settings targeted by HetNV: the data are collected on dense time--energy grids, the signal-to-noise ratio varies across the domain, and the downstream analysis requires repeated smoothing and uncertainty quantification across windows. Exact GP inference or fully latent heteroskedastic GP modeling would be difficult to deploy routinely in this setting.

\subsection{Data structure and modeling assumptions}

For each time index $t$, APS records a current measurement across $B$ discrete energy bins. Let $I_{t,b}$ denote the measured current at time $t$ in energy bin $b$, whose known center energy is $E_b$. Thus, each observation corresponds to a current measurement associated with the time--energy location $(t,E_b)$. In the implementation below, we index observations by the discrete bin label $b$, while the mapping $b \mapsto E_b$ defines the physical energy scale used for interpretation and downstream peak extraction.

We model
\[
  I(t,b) = f(t,b) + \varepsilon(t,b),
  \qquad
  \varepsilon(t,b) \sim N\big(0,\sigma^2(t,b)\big),
\]
where $f$ is a GP over the two-dimensional input $(t,b)$ and $\sigma^2(t,b)$ captures input-dependent observation noise that typically increases in high-energy tails and low-signal regions. Here, $I(t,b)$ denotes the measured instrument current at time $t$ and energy bin $b$.

\subsection{HetNV-based APS processing pipeline and implementation details}

We apply HetNV to estimate a smooth current surface $\widehat{I}(t,b)$ and an input-dependent predictive variance for each time--energy bin. These estimates are incorporated into the APS spacecraft-potential pipeline described in \citet{ulrich2026epee}. 

For each time $t$, the pipeline identifies the energy bin corresponding to the maximum background-subtracted current and evaluates whether this peak is distinguishable from noise. Specifically, a peak is retained if the corresponding 95\% observation-level predictive interval, combining kriging variance and nugget variance, does not include zero.

Under a homoskedastic model, this decision is driven by a single global noise variance, which can lead to inconsistent behavior across the domain: intervals may be overly conservative in low-noise regions and insufficiently adaptive in high-noise regions. This distinction is scientifically important because low-SNR regions often coincide with conditions under which related instruments also experience degraded measurement quality or missing observations, necessitating interpolation or gap-filling procedures. Such regions may nevertheless contain scientifically important structure, including signatures associated with equatorial ionization anomalies or storm-time ionospheric dynamics. By contrast, HetNV replaces the global noise level with a locally estimated variance surface $\hat\sigma^2(t,b)$, allowing the peak-selection rule to adapt to both spatially varying noise and interpolation uncertainty.

Within each APS processing window, time indices and energy-bin indices are first standardized and then rescaled to $[0,1]^2$ prior to neighborhood selection. Unless otherwise stated, HetNV is fit using a Mat\'ern-$3/2$ kernel with neighborhood size $m=20$. The variance surface is initialized using the constant nugget estimate from a homoskedastic Vecchia fit, and alternating mean--variance updates are repeated until either convergence or a maximum of six iterations.

For variance estimation, we smooth the log pseudo-responses using a thin plate spline GAM implemented in \texttt{mgcv}. Unless otherwise stated, we use stabilization parameter $\alpha=0.5$, spline basis dimension $k=60$, shrinkage parameter $\lambda=5$, and convergence tolerance $10^{-3}$ based on the relative change in the estimated variance surface between successive iterations. For larger windows ($n \ge 3000$), smoothing is performed using \texttt{bam()} rather than \texttt{gam()} for computational efficiency. All APS windows are processed using identical model specifications and tuning parameters to support automated large-scale processing without window-specific tuning.

\subsection{Qualitative outcomes}

\begin{figure}[htbp]
    \centering
    \includegraphics[width=\linewidth]{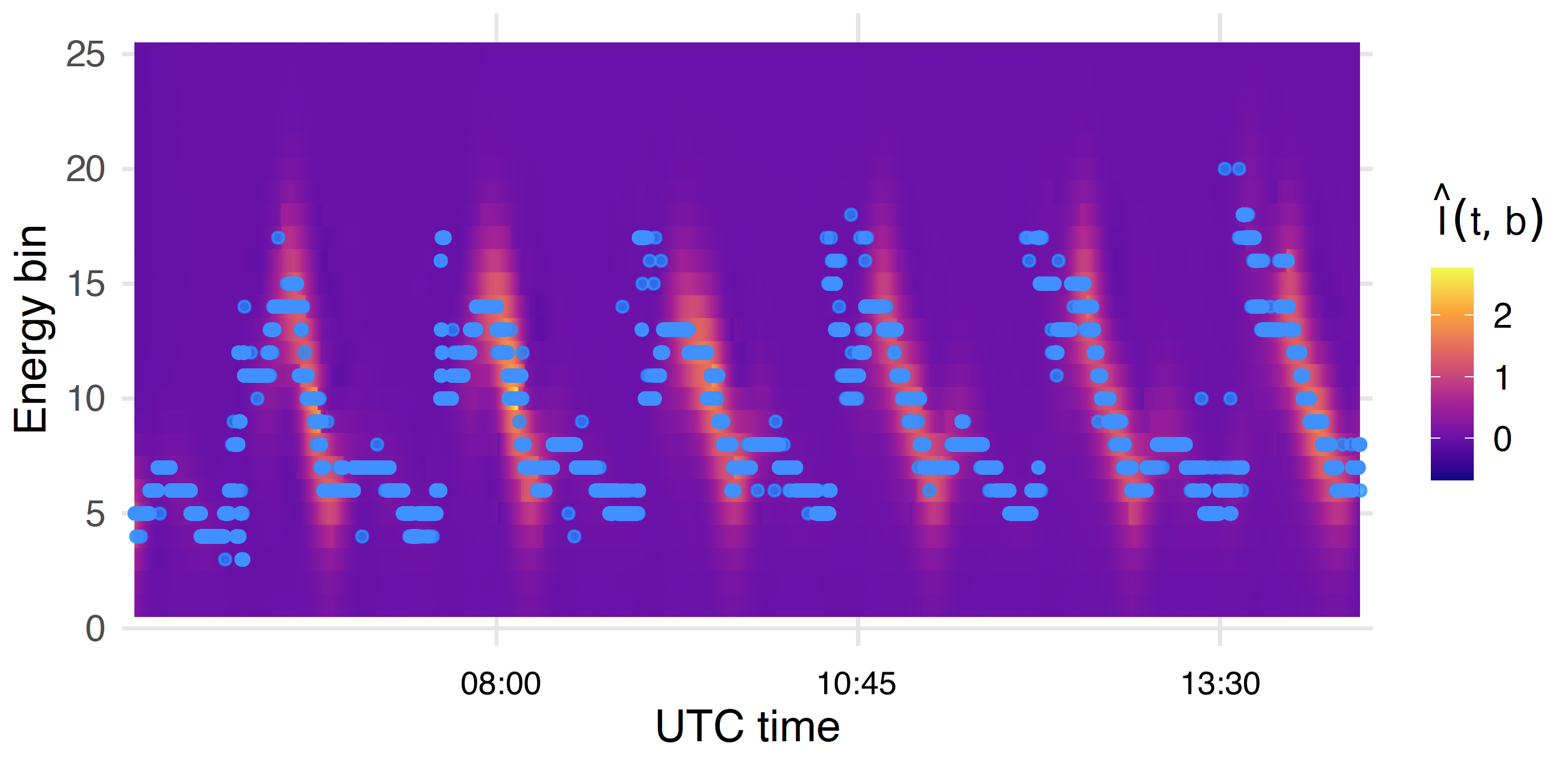}
\caption{HetNV fitted current surface $\hat I(t,b)$ for an approximately 12-hour APS time window from March 19, 2017, with selected peak locations overlaid. At each time $t$, the plotted point denotes the energy bin attaining the maximum of $\hat I(t,b)$. This candidate peak is retained only if its corresponding 95\% predictive interval excludes zero; otherwise, no peak is recorded. For comparison, HomoSV and HetNV select the same peak energy bin at 78.5\% of time points in this window, while differing at 21.5\%. Differences are concentrated in regions where the estimated noise levels diverge across models.}
    \label{fig:aps_peaks}
\end{figure}

\begin{figure}[htbp]
    \centering
    \includegraphics[width=\linewidth]{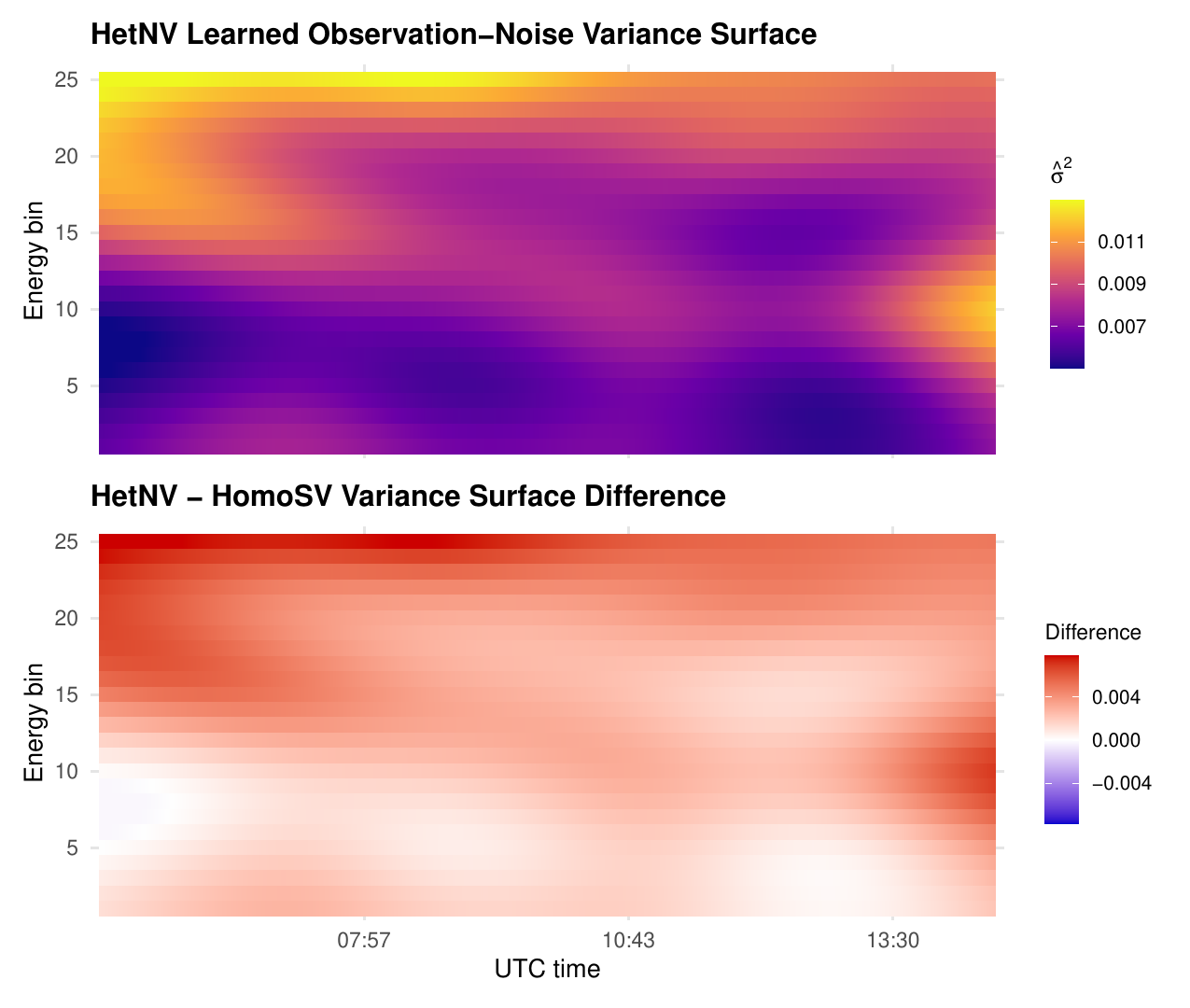}
    \caption{Estimated observation-noise variance surface for the same APS time window shown in \ref{fig:aps_peaks}. 
The top panel shows the HetNV estimate $\hat\sigma^2(t,b)$, while the bottom panel shows its difference relative to the constant variance implied by HomoSV. 
Regions of elevated variance are concentrated in high-energy and low-current regimes, consistent with expected instrument behavior in low signal-to-noise conditions.
These localized increases in estimated variance directly influence the peak-selection rule by widening predictive intervals in noisy regions, reducing the likelihood of spurious peak detections.}
    \label{fig:aps_variance}
\end{figure}

% Talk here about how we are able to better identify signal we suspect is physically true, and therefore allows us to use more of our data. As can be seen in the the above figures, we're able to keep more data poitns that the HetNV model didn't identify as spurious noise, but true signal, which is aligned with what domain experts would expect, while also removing points that we expect to not be physically true or present
We summarize the practical impact of heteroskedastic modeling through three consistent observations across APS windows, focusing on how uncertainty estimation affects downstream signal-detection decisions. Figure~\ref{fig:aps_peaks} illustrates the fitted current surface and peak-selection rule, while Figure~\ref{fig:aps_variance} shows the learned observation-noise variance surface. Because ground-truth peak locations are generally unavailable in in-situ plasma measurements, we interpret differences between HomoSV and HetNV primarily as reflecting changes in uncertainty calibration rather than definitive improvements in detection accuracy. This limitation is not unique to APS, but is characteristic of low-SNR space-plasma measurement settings more broadly, where physically meaningful signals are often difficult to distinguish from instrumental or environmental noise. Consequently, potentially informative observations are frequently discarded because of uncertainty regarding their reliability, making principled uncertainty quantification an important component of downstream scientific analysis and signal-retention decisions.

\emph{(i) Improved handling of low-SNR regions.}
In regions where the underlying current is near the instrument sensitivity, observed traces frequently exhibit isolated high-energy maxima. Under a homoskedastic model, these points can appear spuriously significant because they are evaluated relative to a global noise level. HetNV assigns elevated variance in these regions, resulting in wider predictive intervals and reducing the likelihood that such maxima are incorrectly retained as signal.

\emph{(ii) Spatially adaptive signal detection.}
The use of a predictive distribution to assess whether a peak is distinguishable from zero provides a principled, uncertainty-aware decision rule. With HetNV, this rule adapts locally through $\hat\sigma^2(t,b)$ and the kriging variance, producing peak-selection behavior that reflects both estimated measurement noise and interpolation uncertainty. This contrasts with the homoskedastic case, where a single global threshold implicitly governs all decisions.

\emph{(iii) Structured variance estimates.}
Figure~\ref{fig:aps_variance} shows that the estimated variance surface is strongly structured in both time and energy. Elevated variance is concentrated in high-energy and low-current regions, consistent with known instrument behavior. This structure cannot be captured by a constant-nugget model and directly influences downstream peak selection.

Figure~\ref{fig:aps_peaks} illustrates the resulting differences in selected peaks. Some selected bins appear in regions where the fitted current surface is close to zero because peak selection is performed within each time trace: the plotted bin is the relative maximum across energy bins at that time, not necessarily a large absolute current value. Candidate peaks whose 95\% predictive intervals include zero are treated as indistinguishable from noise and are not retained for downstream analysis. While both models recover similar large-scale signal patterns, discrepancies arise in regions where local uncertainty differs. Peaks retained only by HomoSV tend to occur in regions where a constant variance underestimates noise, while peaks retained only by HetNV reflect regions where local uncertainty is lower than the global baseline.While both models recover similar large-scale signal patterns, discrepancies arise in regions where local uncertainty differs. Peaks retained only by HomoSV tend to occur in regions where a constant variance underestimates noise, while peaks retained only by HetNV reflect regions where local uncertainty is lower than the global baseline.

Taken together, Figures~\ref{fig:aps_peaks} and~\ref{fig:aps_variance} illustrate that heteroskedastic variance estimation affects peak selection both indirectly, through reweighting in the mean fit, and directly, through spatially varying uncertainty thresholds. These results demonstrate that heteroskedastic variance estimation is not only a modeling refinement: it changes which time--energy peaks are retained for downstream spacecraft-potential estimation, thereby affecting the scientific signal passed to later stages of the analysis pipeline.

\section{Discussion}\label{sec:dis}

A key modeling consideration is the distinction between heteroskedasticity and nonstationarity. HetNV models input-dependent observation noise while maintaining a stationary covariance structure for the latent mean process. However, in practice, apparent changes in variability across the input space may arise either from heteroskedastic noise or from nonstationarity in the underlying process (e.g., spatially varying smoothness or correlation range) \citep{Paciorek2003}. These effects can be difficult to disentangle, particularly in small samples or in the absence of replicated observations, since the nugget term may represent both measurement error and unresolved microscale variation \citep{Tang2021}. As a result, some degree of confounding between noise variance and covariance structure is unavoidable in finite samples. HetNV adopts a pragmatic approach by focusing on variance calibration, but extending the framework to jointly model heteroskedasticity and nonstationarity is an important direction for future work. 

HetNV is intentionally designed as a scalable, plug-in alternative to fully latent heteroskedastic GP models. 
The primary predictive accuracy gains of HetNV relative to HomoSV arise when observation noise varies strongly across the input space and when a global nugget cannot calibrate predictive uncertainty.
In such regimes, heteroskedastic reweighting can improve mean-surface estimation and recover spatially varying uncertainty structure, as seen in the 2D experiments and the APS application, although empirical coverage may still be affected by plug-in variance uncertainty and Vecchia approximation error.
Because the mean update relies on a fixed-neighborhood Vecchia approximation, accuracy can degrade when $m$ is too small relative to the effective correlation range or when $n$ grows while $m$ is held fixed.
In practice, increasing $m$ and/or using improved orderings can improve likelihood approximation and coverage.

Because HetNV relies on an alternating plug-in scheme rather than joint posterior inference, it does not provide a full Bayesian posterior over the variance function. This design yields substantial computational benefits but does not fully propagate uncertainty in the estimated variance function into the final predictive uncertainty. The smoothing model (LOESS/GAM), stabilization parameter $\alpha$, and any truncation/shrinkage choices control the bias--variance tradeoff of $\hat{\sigma}^2(\cdot)$; we recommend selecting these defaults once and holding them fixed to support automation.
HetNV targets Gaussian observation noise with input-dependent variance.
When noise is heavy-tailed or contains outliers, additional robustness (e.g., heavier-tailed likelihoods or stronger pseudo-response truncation) may be beneficial.

While HetNV performs well in low- and moderate-dimensional settings, extending the variance model to higher-dimensional inputs may require more structured representations to avoid over-smoothing or instability. Additionally, because the method relies on a plug-in approach, uncertainty in the estimated variance function is not fully propagated into predictive intervals, which can lead to undercoverage in some settings. Finally, performance depends on the choice of smoothing method and stabilization parameters, which may require adjustment in highly irregular or sparse data regimes.

\section{Conclusion}
\label{sec:conclusion}

We have introduced Heteroskedastic Normalized Vecchia (HetNV), a framework for scalable GP regression with input-dependent noise. HetNV combines a Vecchia approximation for the mean function with dimension-adaptive variance estimation based on stabilized residuals. The resulting algorithm alternates between updating the mean GP and smoothing the log-variance, requires no latent-variable inference, and for fixed neighborhood size $m$ and fixed variance-smoother complexity is dominated by $\mathcal{O}(n m^2)$ operations in the sample size $n$.

On synthetic examples, HetNV improves recovery of heteroskedastic uncertainty structure and often improves predictive calibration, while retaining competitive or improved predictive accuracy. The method scales to large datasets, produces interpretable variance surfaces, and avoids the cubic costs associated with dense GP covariance matrices. An application to spacecraft plasma measurements demonstrates that HetNV can support domain-specific pipelines for denoising and peak extraction in large, heteroskedastic datasets.

Future work includes extending HetNV to higher-dimensional input spaces through
structured variance models such as additive or low-rank decompositions, as well
as incorporating nonstationary mean and covariance structures.
An important direction is the integration of replication-based variance
estimators where repeated observations are available, which may further improve
efficiency and stability.
We also intend to develop a dedicated software implementation to facilitate
broader adoption of the proposed methodology, and to pursue follow-up
application-focused studies that examine domain-specific performance in greater
detail.
Beyond the application considered here, we anticipate that HetNV will be useful in
settings such as climate science, environmental monitoring, and large-scale
simulation studies, where scalable modeling and reliable heteroskedastic
uncertainty quantification are critical.

\section*{Software and Reproducibility}

A software implementation of the HetNV methodology is currently under development and will be released as an R package following publication. The package will include functions for model fitting, prediction, and visualization, along with scripts to reproduce the simulation studies presented in this paper. The code will be made publicly available upon institutional approval for release at: https://github.com/kydpotter/HetNV.

All simulation settings and modeling choices are fully specified in the manuscript to ensure reproducibility. Additional implementation details are available from the authors upon request.

\section*{Acknowledgments}

Research presented in this manuscript was supported by the Laboratory Directed Research and Development program of Los Alamos National Laboratory under project number 20240045DR. D.C.S. is supported by an NSERC Discovery Grant (RGPIN-2021-03985). D.B. acknowledges partial funding support through NSERC Discovery Grants.

This research used resources provided by the Darwin testbed at Los Alamos National Laboratory (LANL) which is funded by the Computational Systems and Software Environments subprogram of LANL's Advanced Simulation and Computing program (NNSA/DOE).

{Approved for public release: LA-UR-26-24130}.

\bibliographystyle{apalike}
\bibliography{refs}
\newpage
\appendix
% \documentclass[11pt]{article}

% % =========================================================
% % Packages
% % =========================================================
% \usepackage[margin=1in]{geometry}
% \usepackage{amsmath,amssymb}
% \usepackage{graphicx}
% \usepackage{float}
% \usepackage{caption}
% \usepackage{subcaption}
% \usepackage{setspace}

% =========================================================
% Formatting
% =========================================================
% \onehalfspacing

% % Supplementary figure numbering: S1, S2, ...
% \renewcommand{\thefigure}{S\arabic{figure}}

% % =========================================================
% % Document
% % =========================================================
% % \begin{document}

\begin{center}
{\Large\bfseries Supplementary Material}\\[0.75em]

{\large
Scalable Heteroskedastic Gaussian Process Models\\
for Large Inhomogeneous Datasets
}
\end{center}

\vspace{1em}

\section*{Additional Figures and Diagnostics}
\label{sec:supp-additional-figs}

This supplementary material provides additional visual diagnostics for the
two-dimensional heteroskedastic simulation, including truth surfaces, fitted
mean and noise surfaces, and replicate-level simulation diagnostics. We also
include boxplots summarizing the distributions of empirical 95\% coverage,
runtime, mean-estimation error, and predictive-standard-deviation recovery
across Monte Carlo replicates.

% =========================================================
% Truth surfaces
% =========================================================

\begin{figure}[p]
  \centering
  \includegraphics[width=0.85\linewidth]
  {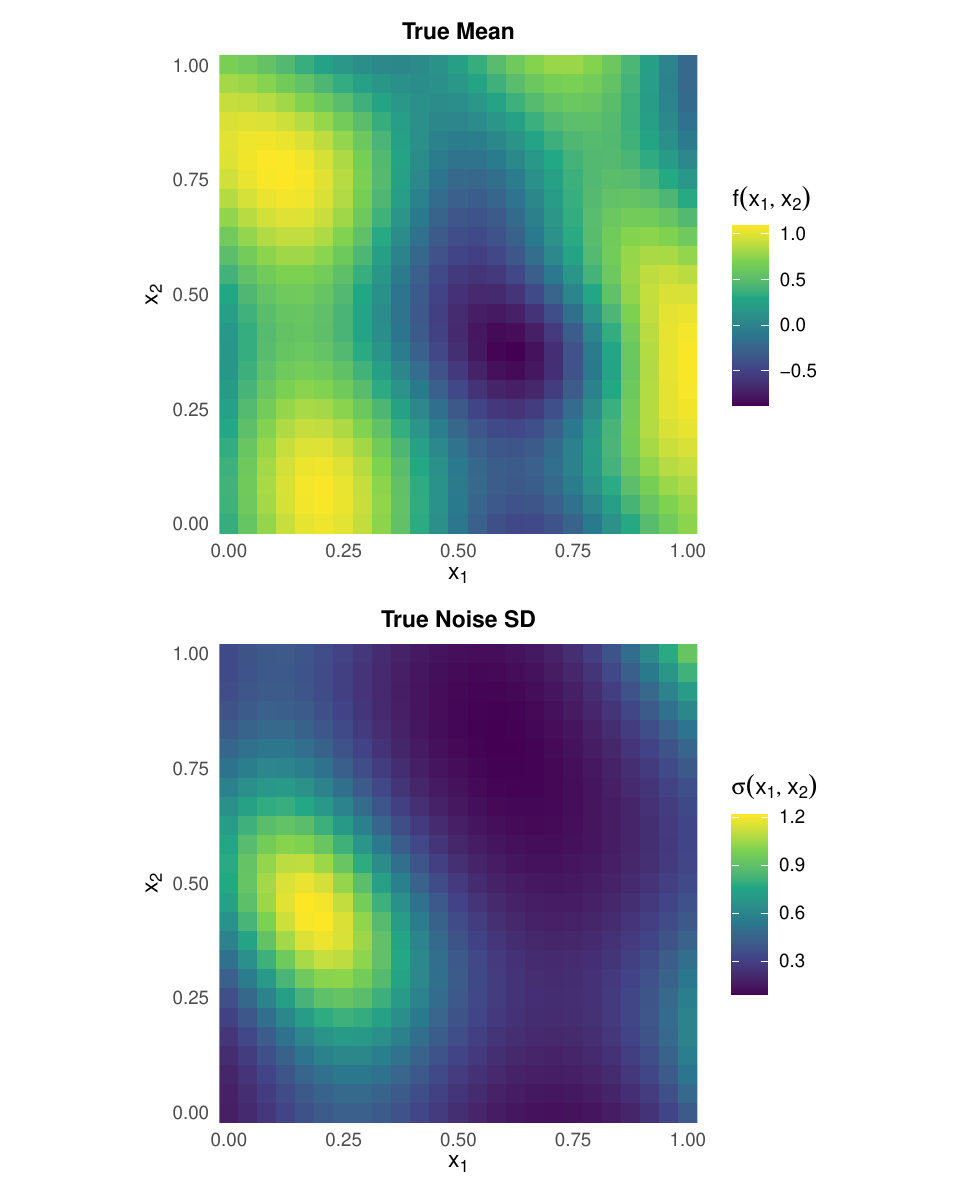}

  \caption{
  True mean surface $f(x_1,x_2)$ and true noise standard deviation surface
  $\sigma(x_1,x_2)$ for the two-dimensional heteroskedastic simulation.
  }
  \label{fig:app_true_mean}
\end{figure}

% =========================================================
% Fitted surfaces
% =========================================================

\begin{figure}[p]
  \centering

  \begin{subfigure}{\textwidth}
    \centering
    \includegraphics[width=0.85\linewidth]
    {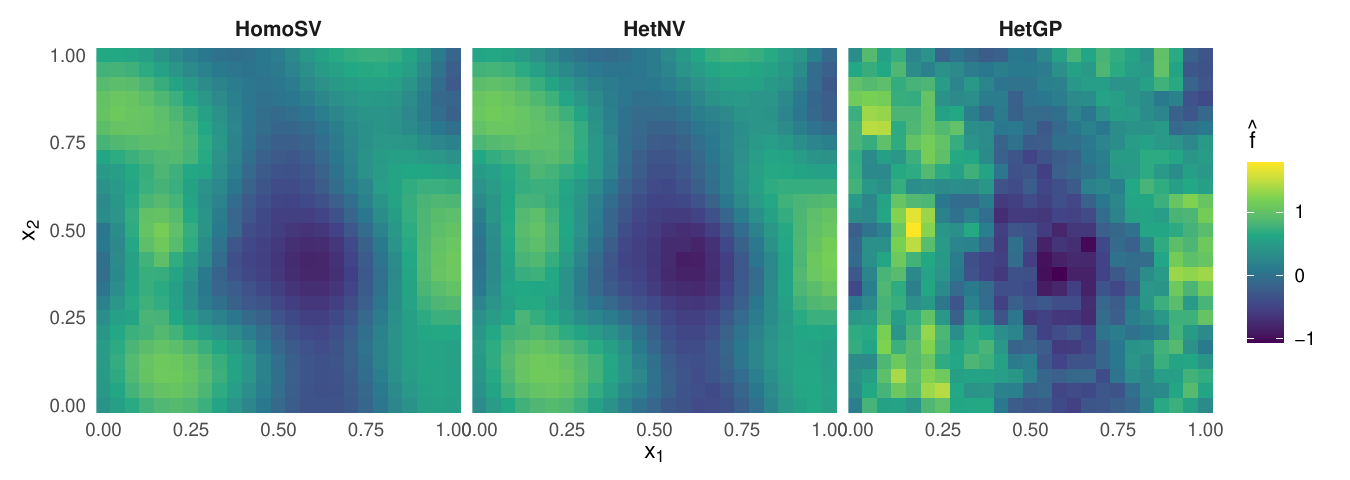}

    \caption{
    Fitted mean surfaces $\hat f(x_1,x_2)$ for all models.
    }
    \label{fig:app_mean_fits}
  \end{subfigure}

  \vspace{1em}

  \begin{subfigure}{\textwidth}
    \centering
    \includegraphics[width=0.85\linewidth]
    {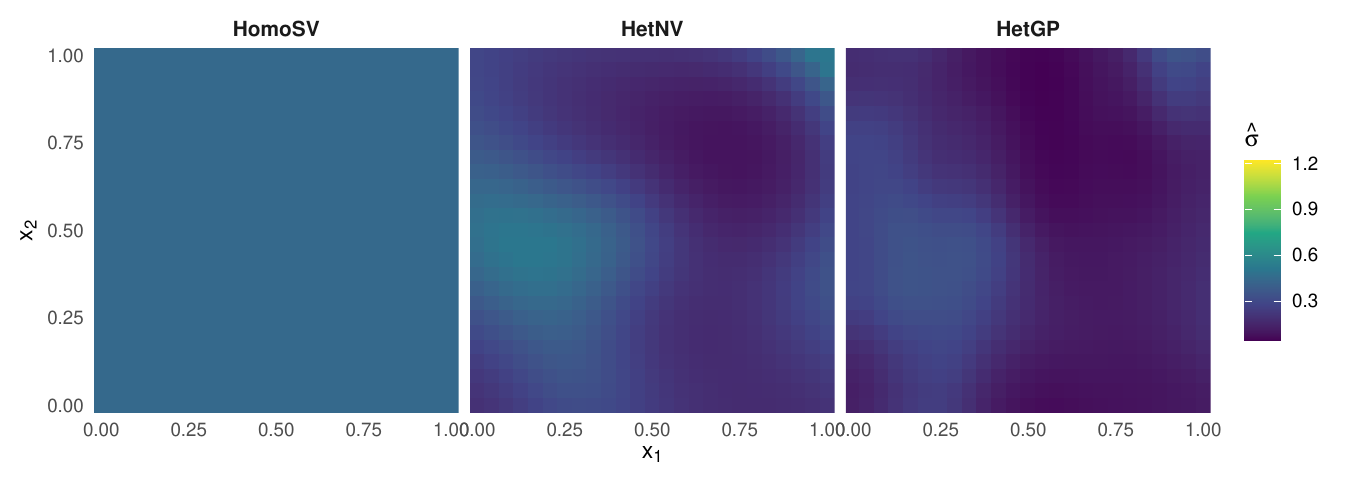}

    \caption{
    Fitted noise standard deviation surfaces
    $\hat\sigma(x_1,x_2)$ for all models.
    }
    \label{fig:app_noise_fits}
  \end{subfigure}

  \caption{
  Estimated mean and noise standard deviation surfaces for all fitted models
  in the two-dimensional heteroskedastic simulation.
  Top: fitted mean surfaces $\hat f(x_1,x_2)$.
  Bottom: fitted noise standard deviation surfaces
  $\hat\sigma(x_1,x_2)$.
  }
  \label{fig:app_2d_fitted_surfaces}
\end{figure}

% =========================================================
% Replicate-level diagnostics
% =========================================================

\clearpage

\section*{Replicate-Level Simulation Diagnostics}

\begin{figure}[!ht]
\centering
\includegraphics[width=0.95\textwidth]
{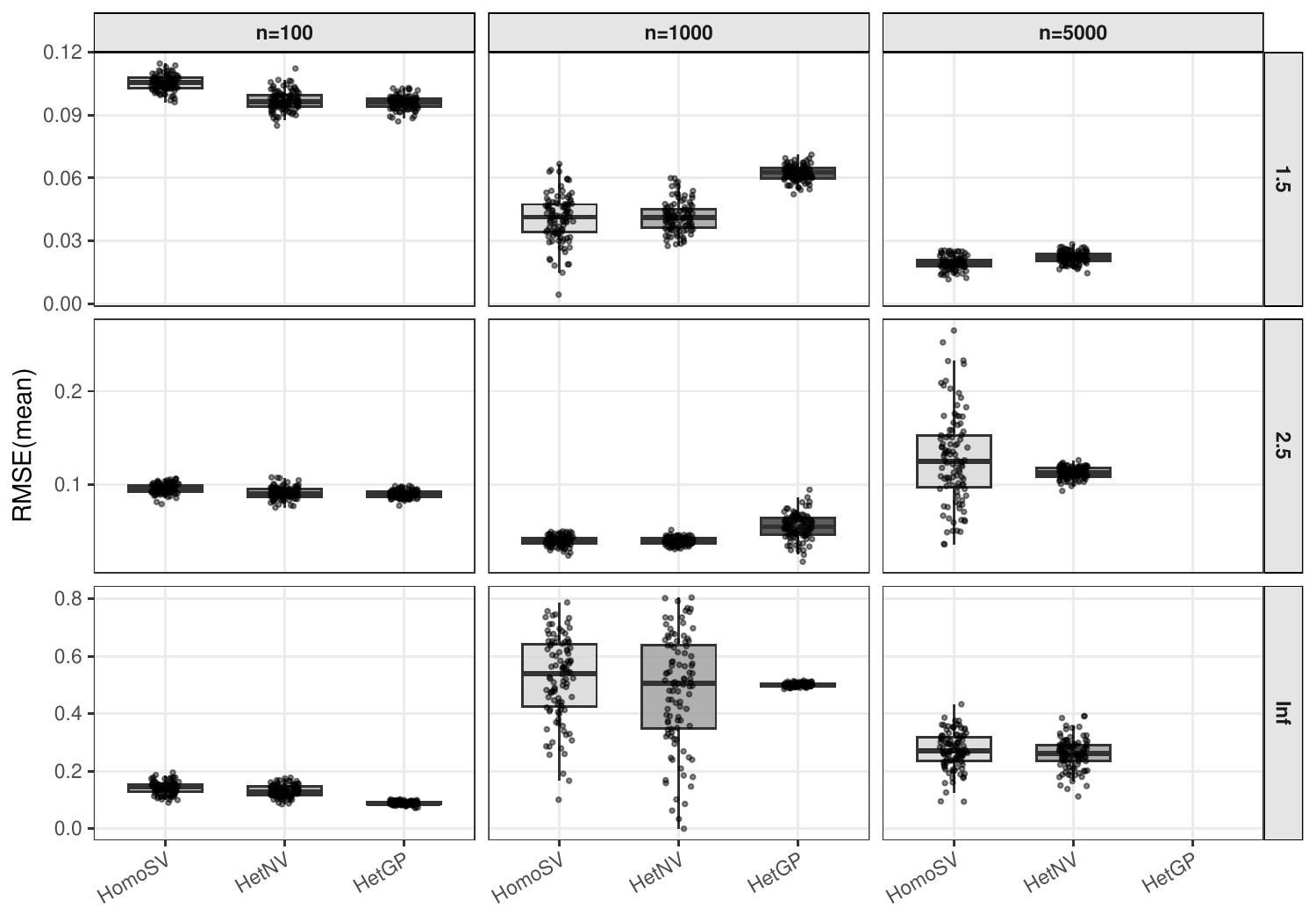}
\caption{
RMSE of the latent mean estimate across Monte Carlo replicates for the
one-dimensional heteroskedastic setting. Boxes summarize
replicate-to-replicate variability, and points show individual simulated
replicates.
}
\label{fig:app-1d-rmse-boxplot}
\end{figure}

\begin{figure}[!ht]
\centering
\includegraphics[width=0.95\textwidth]
{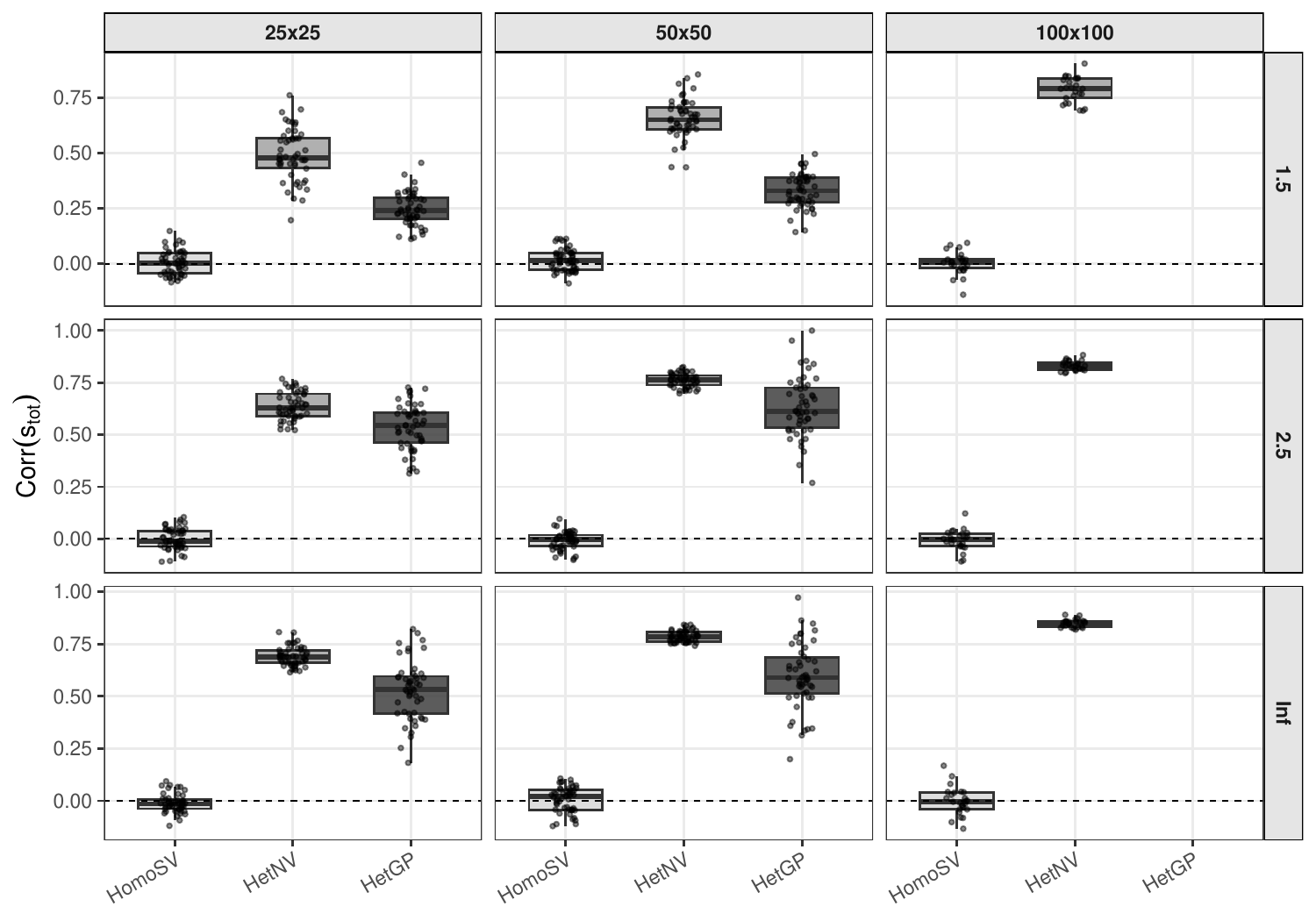}
\caption{
Correlation between estimated and true total predictive standard deviation
across Monte Carlo replicates for the two-dimensional heteroskedastic setting.
The dashed line marks zero correlation. HetNV consistently improves recovery
of the spatially varying predictive standard deviation surface relative to
HomoSV.
}
\label{fig:app-2d-sd-corr-boxplot}
\end{figure}

Figures~S\ref{fig:app-1d-rmse-boxplot}--S\ref{fig:app-2d-sd-corr-boxplot}
show replicate-level distributions for the primary simulation metrics.
These plots provide additional information about Monte Carlo variability
around the summary statistics reported in the main manuscript.

% -------------------------
% 1D diagnostics
% -------------------------

\begin{figure}[!ht]
\centering
\includegraphics[width=0.95\textwidth]
{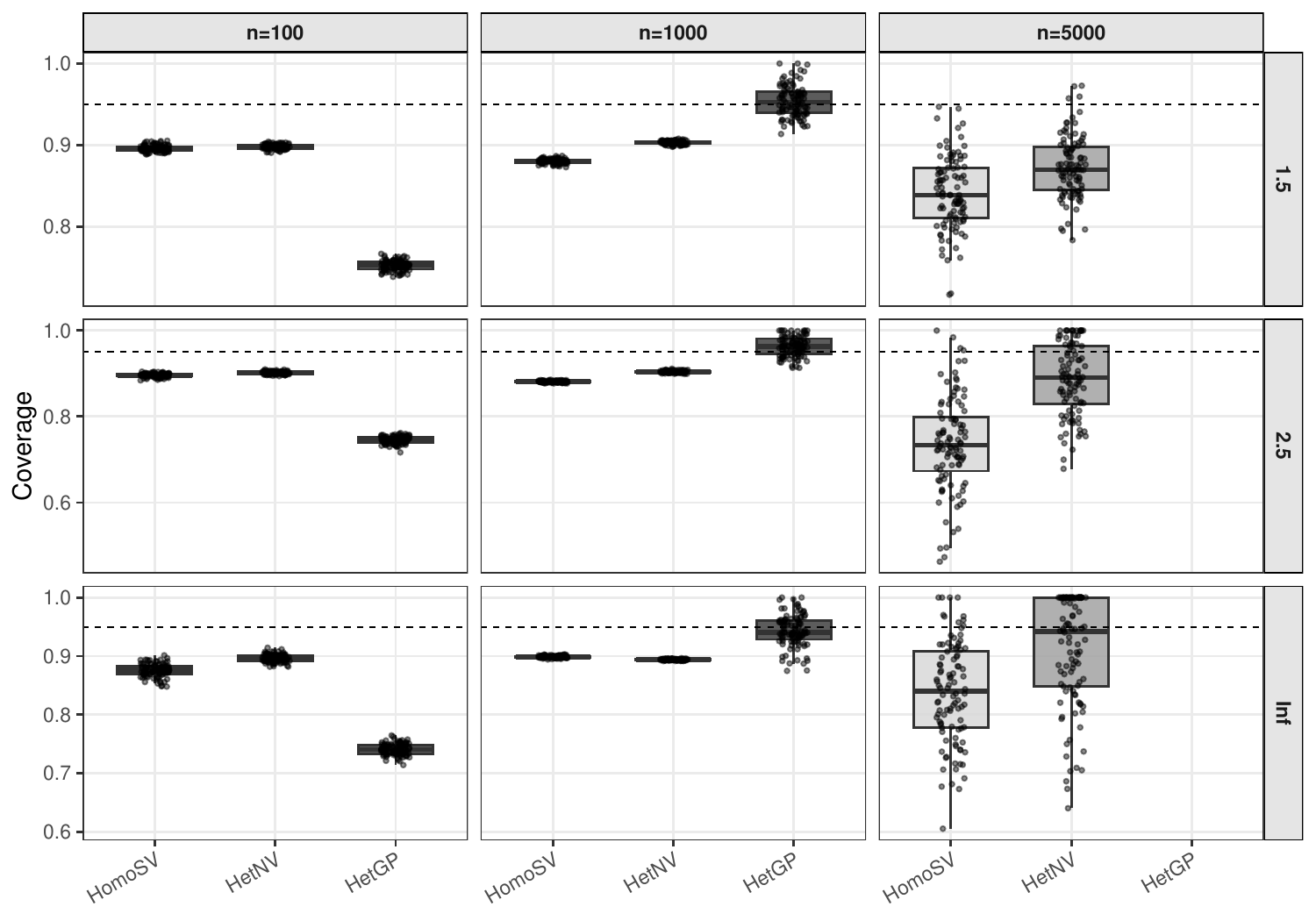}
\caption{
Empirical 95\% observation-level coverage across Monte Carlo replicates
for the one-dimensional heteroskedastic setting. The dashed line denotes
nominal 95\% coverage.
}
\label{fig:app-1d-coverage-boxplot}
\end{figure}

\begin{figure}[!ht]
\centering
\includegraphics[width=0.95\textwidth]
{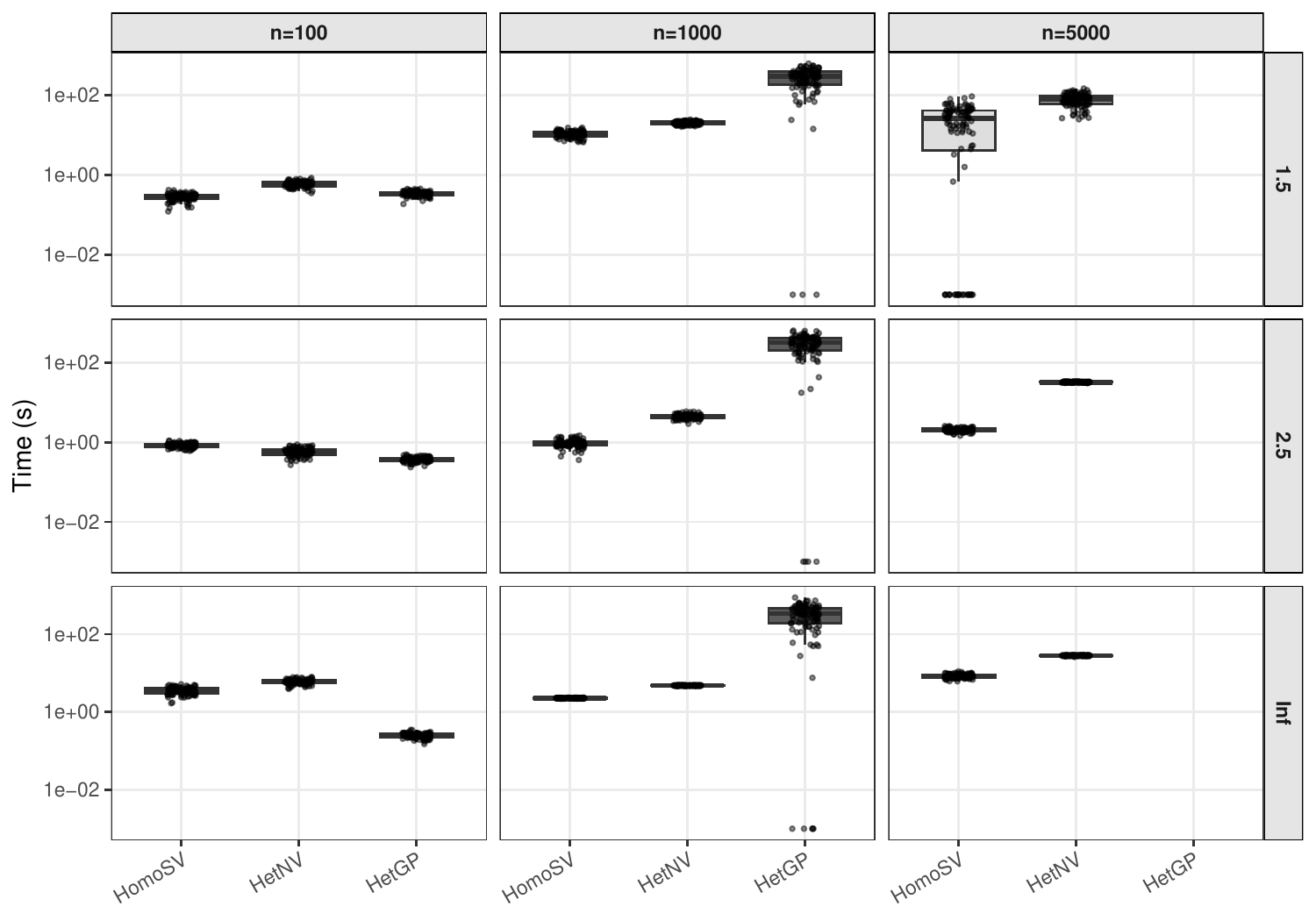}
\caption{
Runtime in seconds across Monte Carlo replicates for the one-dimensional
heteroskedastic setting, shown on a log scale.
}
\label{fig:app-1d-time-boxplot}
\end{figure}

\begin{figure}[!ht]
\centering
\includegraphics[width=0.95\textwidth]
{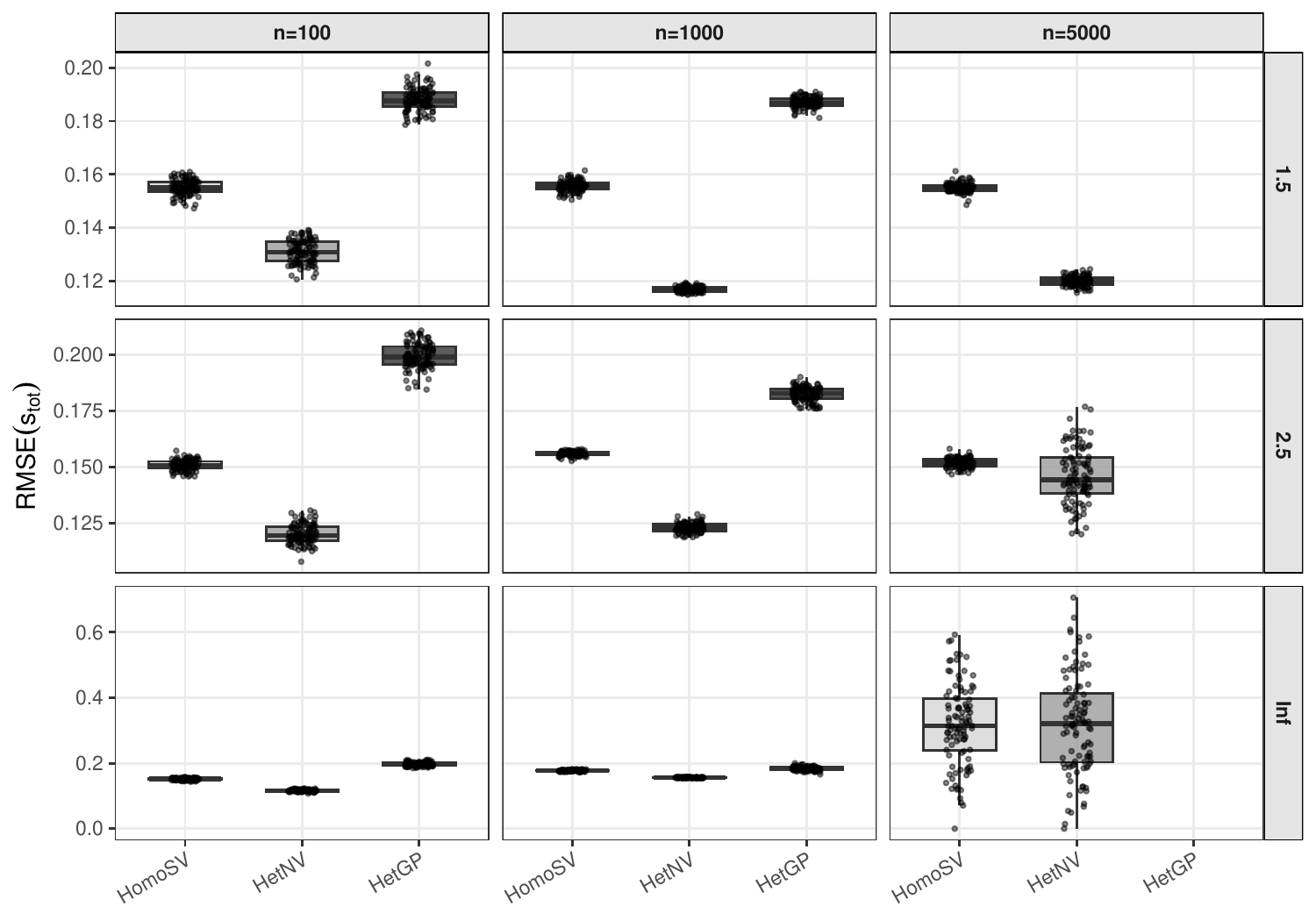}
\caption{
RMSE of the total predictive standard deviation surface across Monte Carlo
replicates for the one-dimensional heteroskedastic setting.
}
\label{fig:app-1d-sd-rmse-boxplot}
\end{figure}

\begin{figure}[!ht]
\centering
\includegraphics[width=0.95\textwidth]
{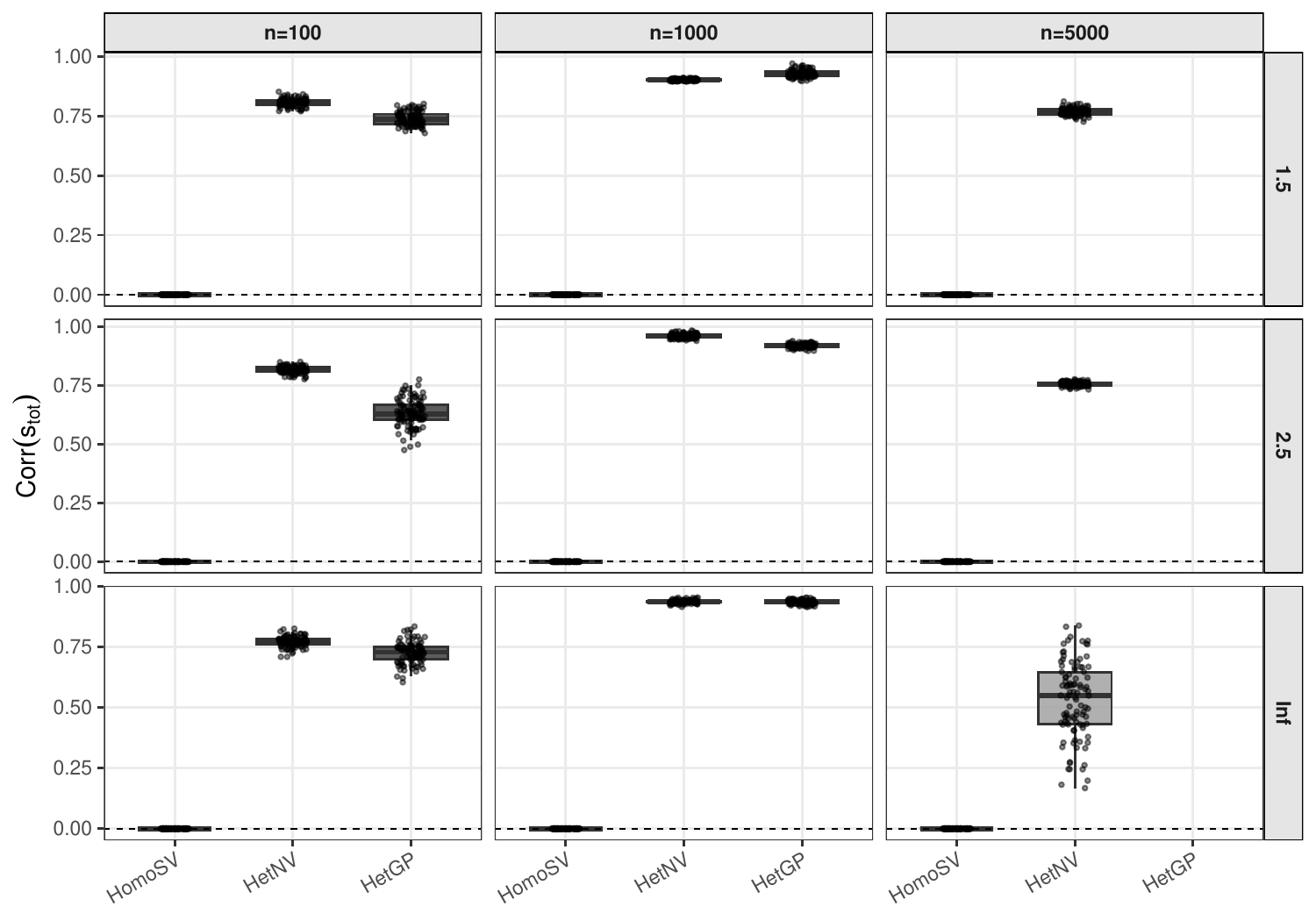}
\caption{
Correlation between estimated and true total predictive standard deviation
across Monte Carlo replicates for the one-dimensional heteroskedastic setting.
The dashed line marks zero correlation.
}
\label{fig:app-1d-sd-corr-boxplot}
\end{figure}

% -------------------------
% 2D diagnostics
% -------------------------

\begin{figure}[!ht]
\centering
\includegraphics[width=0.95\textwidth]
{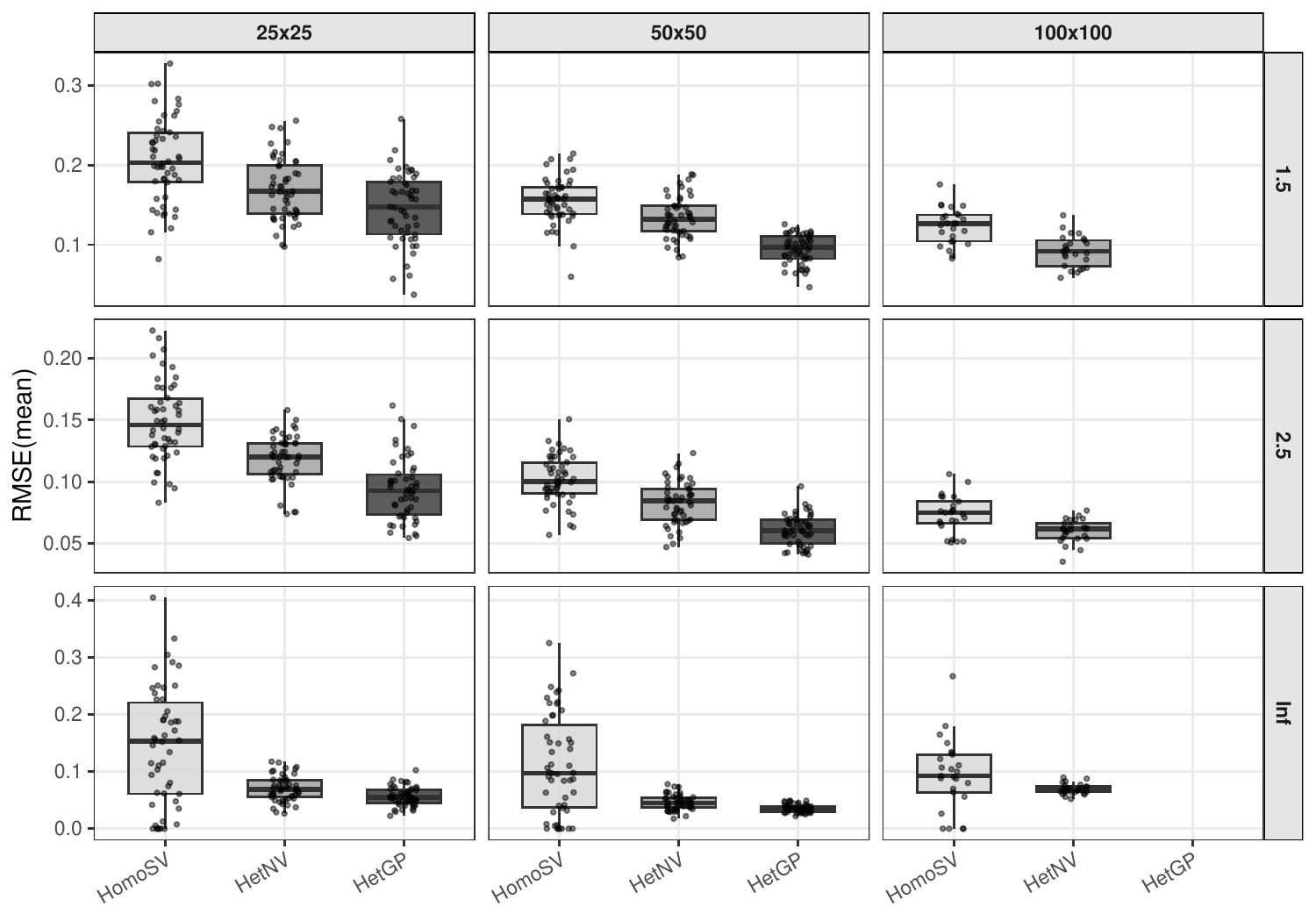}
\caption{
RMSE of the latent mean estimate across Monte Carlo replicates for the
two-dimensional heteroskedastic setting. Columns correspond to grid
resolution and rows correspond to Mat\'ern smoothness.
}
\label{fig:app-2d-rmse-boxplot}
\end{figure}

\begin{figure}[!ht]
\centering
\includegraphics[width=0.95\textwidth]
{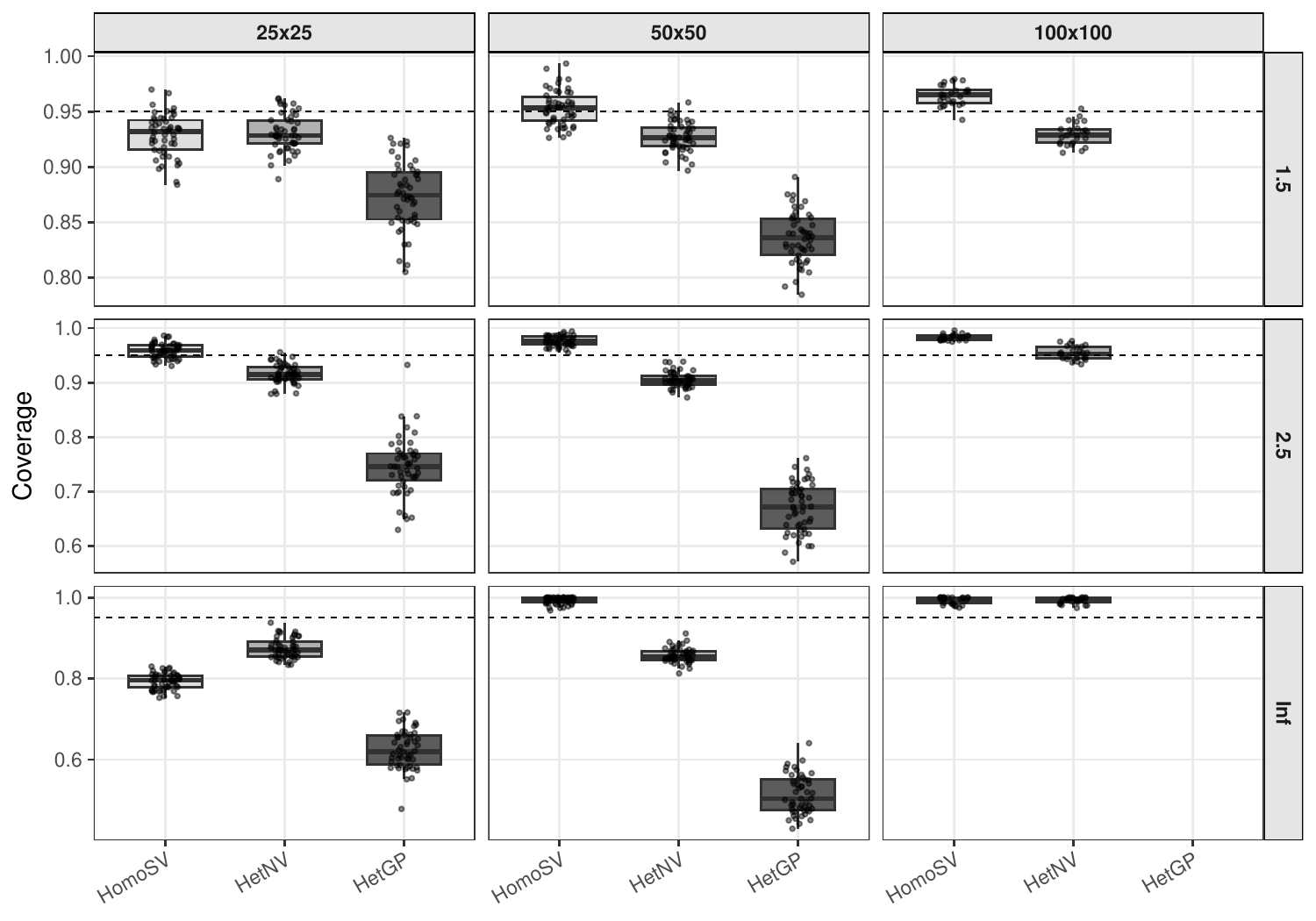}
\caption{
Empirical 95\% observation-level coverage across Monte Carlo replicates
for the two-dimensional heteroskedastic setting. The dashed line denotes
nominal 95\% coverage.
}
\label{fig:app-2d-coverage-boxplot}
\end{figure}

\begin{figure}[!ht]
\centering
\includegraphics[width=0.95\textwidth]
{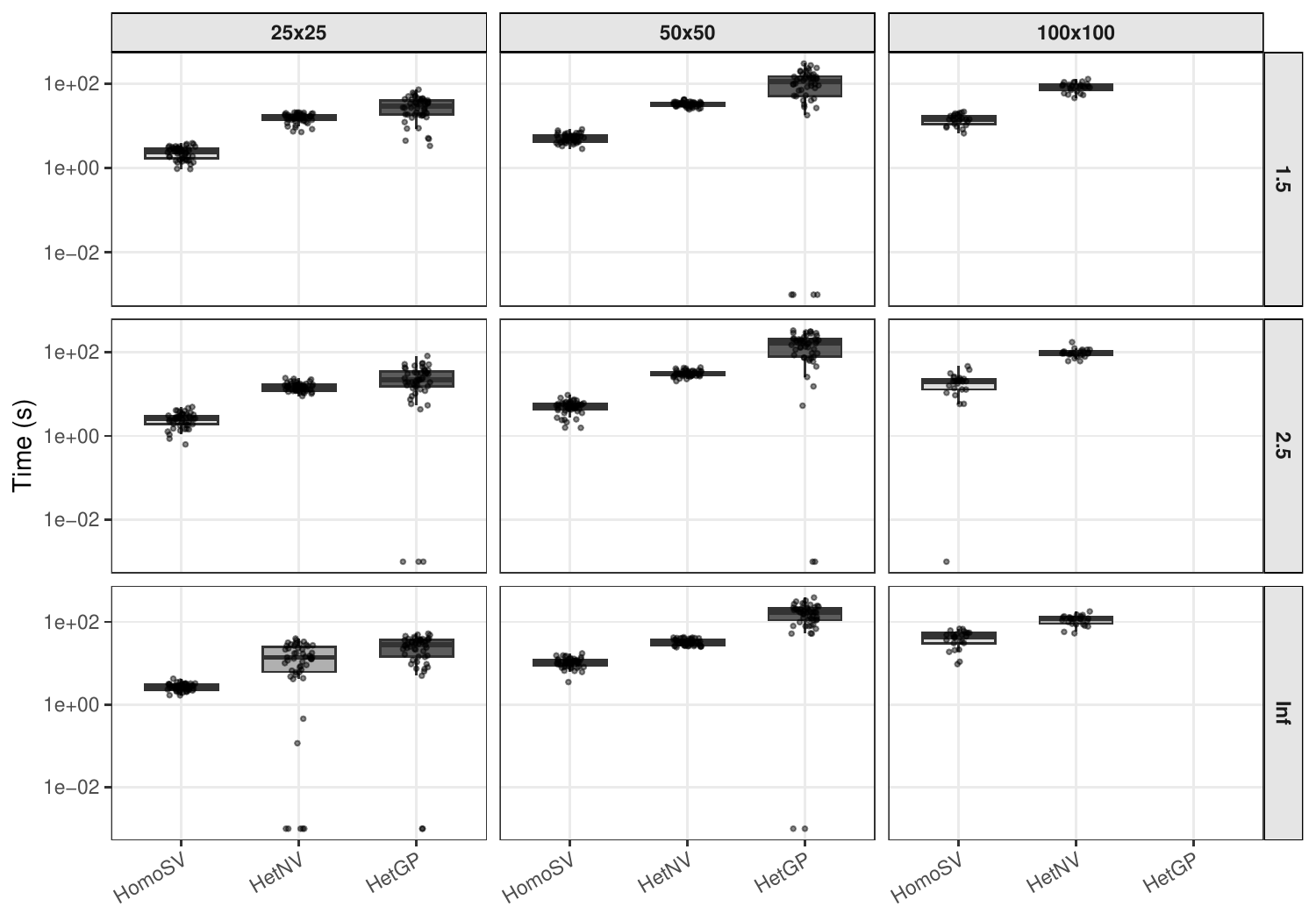}
\caption{
Runtime in seconds across Monte Carlo replicates for the two-dimensional
heteroskedastic setting, shown on a log scale.
}
\label{fig:app-2d-time-boxplot}
\end{figure}

\begin{figure}[!ht]
\centering
\includegraphics[width=0.95\textwidth]
{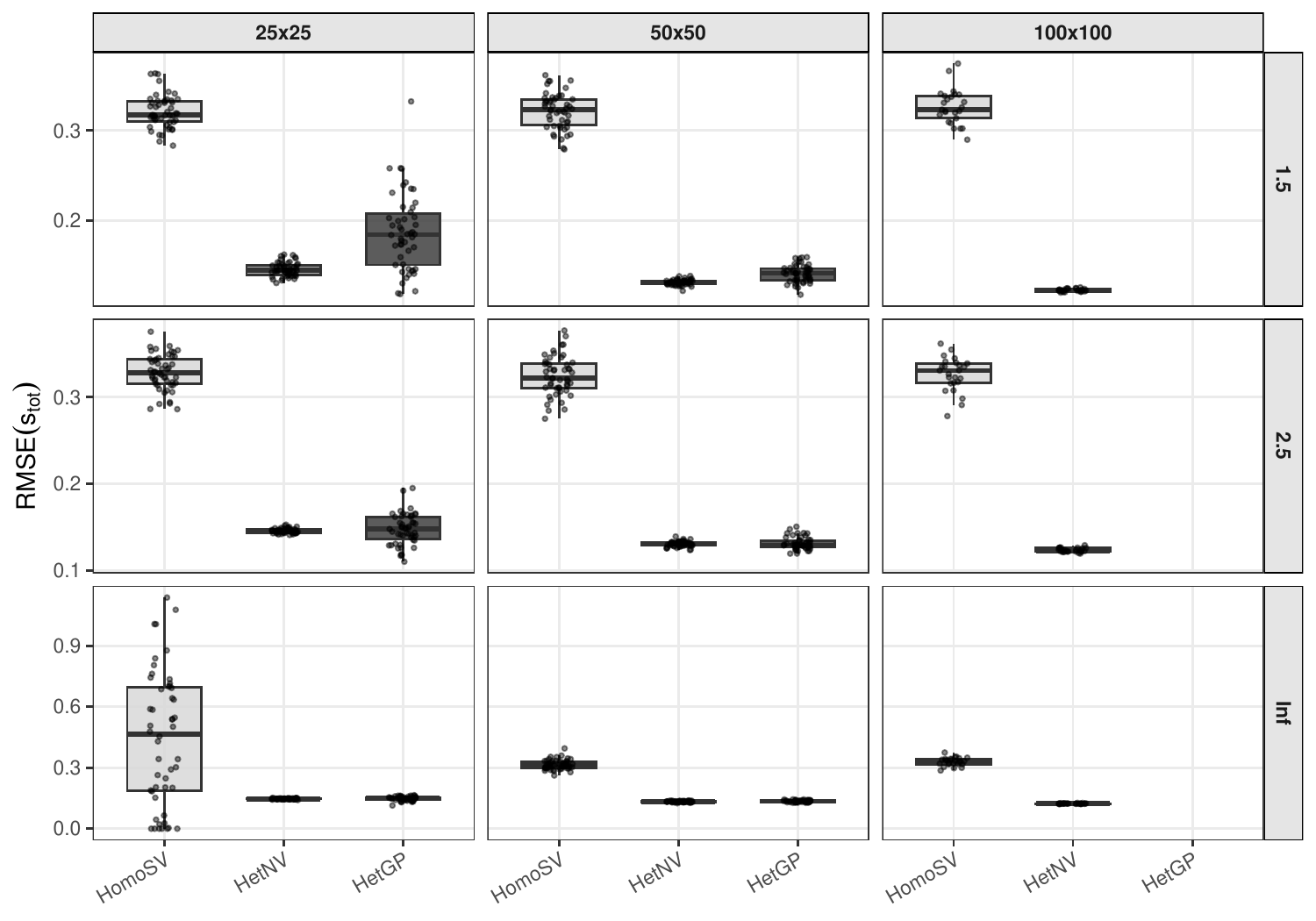}
\caption{
RMSE of the total predictive standard deviation surface across Monte Carlo
replicates for the two-dimensional heteroskedastic setting.
}
\label{fig:app-2d-sd-rmse-boxplot}
\end{figure}

\end{document}